\documentclass{aa} % for a referee version
\usepackage{arydshln}
\usepackage[toc,page]{appendix}
\usepackage[colorlinks = true,
            linkcolor = blue,
            urlcolor  = blue,
            citecolor = blue,
            anchorcolor = blue]{hyperref}
\usepackage{colortbl}
\usepackage[dvipsnames]{xcolor}
\usepackage[none]{hyphenat}
\usepackage{graphicx}
\usepackage{txfonts}
\usepackage{hyperref}
\usepackage{multirow}
\usepackage{enumitem}
\usepackage{placeins}
\usepackage{float}
\DeclareRobustCommand{\ion}[2]{%
\relax\ifmmode
\ifx\testbx\f@series
{\mathbf{#1\,\mathsc{#2}}}\else
{\mathrm{#1\,\mathsc{#2}}}\fi
\else\textup{#1\,{\mdseries\textsc{#2}}}%
\fi}
\begin{document}

   \title{CO-CAVITY pilot survey: Molecular gas and star formation in void galaxies
   \thanks{The CO spectra are available in electronic form at the CDS via anonymous ftp to cdsarc.u-strasbg.fr (130.79.128.5) or via http://cdsweb.u-strasbg.fr/cgi-bin/qcat?J/A+A/}
}

   \subtitle{}

   \author{J.\,Domínguez-Gómez\inst{1} 
          \and
          U.\,Lisenfeld\inst{1,2}
          \and
          I.\,Pérez\inst{1,2}
          \and
          Á. R.\,López-Sánchez\inst{3,4,5,6}
          \and
          S.\,Duarte Puertas \inst{7,8}
          \and
          J.\,Falcón-Barroso\inst{9,10}
          \and
          K.\,Kreckel\inst{11}
          \and
          R. F.\,Peletier\inst{12}
          \and
          T.\,Ruiz-Lara\inst{12}
          \and
          R.\,van\,de\,Weygaert\inst{12}
          \and
          J. M.\,van\,der\,Hulst\inst{12}
          \and
          S.\,Verley\inst{1,2}
          }
   \institute{
   Universidad de Granada (ugr), Departamento de Física Teórica y del Cosmos, Campus Fuente Nueva, Edificio Mecenas, 18071, Granada, Spain. \email{jesusdg@ugr.es, ute@ugr.es, isa@ugr.es} 
   \and
   Instituto Carlos I de Física Tórica y Computacional, Facultad de Ciencias,
 E-18071 Granada, Spain
   \and
   Australian Astronomical Optics, Macquarie University, 105 Delhi Rd, North Ryde, NSW 2113, Australia.
   \and
   Department of Physics and Astronomy, Macquarie University, NSW 2109, Australia.
   \and
   Macquarie University Research Centre for Astronomy, Astrophysics \& Astrophotonics, Sydney, NSW 2109, Australia.
   \and
   ARC Centre of Excellence for All Sky Astrophysics in 3 Dimensions (ASTRO-3D), Australia.
   \and
   Département de Physique, de Génie Physique et d’Optique, Université Laval, and Centre de Recherche en Astrophysique du Québec (CRAQ), Québec, QC, G1V 0A6, Canada
   \and
   Instituto de Astrofísica de Andalucía - CSIC, Glorieta de la Astronomía s.n., 18008 Granada, Spain
   \and
   Instituto de Astrofísica de Canarias, Vía Láctea s/n, 38205 La Laguna, Tenerife, Spain.
   \and
   Departamento de Astrofísica, Universidad de La Laguna, 38200 La Laguna, Tenerife, Spain.
   \and
   Astronomisches Rechen-Institut, Zentrum für Astronomie der Universität Heidelberg, Mönchhofstraße 12-14, D-69120 Heidelberg, Germany.
   \and
   Kapteyn Astronomical Institute, University of Groningen, Landleven 12, 9747 AD Groningen, The Netherlands.
   }

   \date{Received September 15, 1996; accepted March 16, 1997}

% \abstract{}{}{}{}{} 
% 5 {} token are mandatory
 
  \abstract
   {Voids are the most under-dense large-scale regions in the Universe. Galaxies inhabiting voids are one of the keys for understanding the intrinsic processes of galaxy evolution, as external factors such as multiple galaxy mergers or a dense self-collapsing environment are negligible.
   }
   {We present the first molecular gas mass survey  of void galaxies.  We compare these new data together with data for the atomic gas mass ($M_{\rm H{\scriptscriptstyle  I}}$)  and star formation rate (SFR) from the literature to those of  galaxies in filaments and walls in order to better understand how molecular gas and star formation are related to the large-scale environment.
   }  
   {We observed at the IRAM 30 m telescope the $\rm CO(1-0)$ and $\rm CO(2-1)$ emission of 20 void galaxies selected from the Void Galaxy Survey (VGS), with a stellar mass range from $\rm 10^{8.5}$ to $\rm 10^{10.3}M_{\odot}$. We detected 15 objects in at least one CO line. We compared the molecular gas mass ($M_{\rm H_2}$), the star formation efficiency (SFE $\rm=SFR/M_{\rm H_2}$), the atomic gas mass, the molecular-to-atomic gas mass ratio, and the specific star formation rate (sSFR) of the void galaxies with two control samples of galaxies in filaments and walls, selected from xCOLD GASS and EDGE-CALIFA, for different stellar mass bins and taking the star formation activity into account.
   }
   {In general, we do not find any significant differences between void galaxies and the control sample. In particular, we do not find any evidence for a difference in the molecular gas mass or molecular gas mass fraction. For the other parameters (SFE, atomic gas mass, molecular-to-atomic gas mass ratio, and sSFR), we also find similar (within the errors) mean values between void and filament and wall galaxies when the sample is limited to star-forming galaxies. We find no evidence for an enhanced sSFR in void galaxies. Some tentative differences emerge when trends with stellar mass are studied: The SFE of void galaxies might be lower than in filament and wall galaxies for low stellar masses, and there might be a trend of increasing deficiency in the H{\tiny I} content in void galaxies compared to galaxies in filaments and walls for higher stellar masses, accompanied by an increase in the molecular-to-atomic gas mass ratio. However, all trends with stellar mass  are based on a low number of galaxies and need to be confirmed for a larger sample.
   } 
%  conclusions heading (optional), leave it empty if necessary 
   {The results for the molecular gas mass for a sample of 20 voids galaxies allowed us to make a statistical comparison to galaxies in filaments and walls for the first time. We do not find any significant differences of the molecular gas properties and the SFE, but we note that a larger sample is necessary to confirm this and be sensitive to subtle trends.
   }


   \maketitle
%
%-------------------------------------------------------------------
\section{Introduction}
    \label{sec:intro}

    Large redshift galaxy surveys such as the Sloan Digital Sky Survey \citep[SDSS, ][]{2000AJ....120.1579Y}, the 2dF Galaxy Redshift Survey \citep[2dFGRS, ][]{2001MNRAS.328.1039C}, or the 2MASS Redshift Survey \citep[2MRS, ][]{2012ApJS..199...26H} have shown that galaxies are not uniformly distributed in the Universe, but form a hierarchical structure of filaments and walls. This structure surrounds the voids, which are vast regions (10-30 $\rm h^{-1}$ Mpc in diameter) that are almost devoid of galaxies \citep{2001ApJ...557..495P, 2011AJ....141....4K, 2012MNRAS.421..926P, 2012ApJ...744...82V, 2016IAUS..308..493V}. Voids are the most under-dense environments in the Universe, and they are inhabited by the void galaxy population, which is partially distributed along filament-like substructures throughout. These substructures have been confirmed and were modelled by numerical simulations \citep{1993MNRAS.263..481V, 2004MNRAS.350..517S, 2013MNRAS.428.3409A, 2013MNRAS.434.1192R, 2013MNRAS.435..222R} and were also observed \citep{2006MNRAS.372.1710P, 2013AJ....145..120B, 2014MNRAS.440L.106A}. The void galaxies are negligibly affected by multiple galaxy mergers or a dense self-collapsing environment, so they represent a unique galaxy population based on which can be unveiled how the large-scale environment affects the assembly, evolution, and properties of galaxies. 

    Previous studies have shown that void galaxies have bluer colours, lower stellar masses ($M_\star$), and later morphological types on average than galaxies in filaments and walls \citep{2004ApJ...617...50R, 2007ApJ...658..898P, 2012MNRAS.426.3041H, 2012AJ....144...16K, 2021ApJ...906...97F}. The fraction of galaxies with an elevated SFR is higher in voids than in denser environments, but no difference is found when the morphology, luminosity, or $M_\star$ is fixed \citep{2006MNRAS.372.1710P,2012AJ....144...16K, 2014MNRAS.445.4045R}. However, other studies found that void galaxies have an enhanced SFR for a given $M_\star$ or luminosity \citep{2005ApJ...624..571R, 2016MNRAS.458..394B, 2021ApJ...906...97F}. The atomic gas (H{\tiny I}) properties of void galaxies are similar to galaxies in filaments and walls \citep{1996AJ....111.2150S, 2012AJ....144...16K}. Together with the fact that the  small-scale clustering in the voids (within a projected distance of 600 kpc and \makebox{200 km $\rm s^{-1}$}) is also similar to what is found in filaments and walls, this suggests that the small-scale rather than the large-scale environment  of the galaxies affects their gas content and evolution \citep{1996AJ....111.2150S, 2012AJ....144...16K}. However, a recent study has found that void galaxies have higher atomic gas masses than galaxies in filaments and walls \citep{2021ApJ...906...97F}. These discrepancies suggest that further study is needed of how star formation progresses within void galaxies.

    As star formation is strongly regulated by the presence (or absence) of molecular gas \citep{2008AJ....136.2782L, 2011ApJ...730L..13B}, the molecular gas mass is a crucial parameter that allows us to quantify the potential for future star formation. Furthermore, by comparing the molecular gas mass to  the SFR, the stellar mass, and the atomic gas mass, we can search for  possible differences in the star formation process within void galaxies compared to galaxies in filaments and walls. 

    Currently, there is no statistically significant sample of void galaxies with $\rm H_2$ data. Only three previous studies have analysed the $\rm H_2$ content of void galaxies \citep{1997AJ....114.1753S, 2013AJ....145..120B, 2015ApJ...815...40D}, and they suggested that the $\rm H_2$ properties are similar to galaxies in filaments and walls. However, these results are based on very few galaxies, between one to five objects in each study, which is not enough to draw any statistical conclusions. More statistics is needed to better understand the process of star formation and the properties of the molecular gas phase in void galaxies.

    In particular, we need to compare the molecular-to-atomic gas mass ratio and the star formation efficiency (${\rm SFE=SFR}/M_{\rm H_2}$) in void galaxies to those galaxies in filaments and walls in order to better understand how the processes of molecular gas consumption and star formation are related to the large-scale environment. Furthermore, we need to measure the molecular gas mass ($M_{\rm H_2}$) in to order obtain the total neutral gas ($M_{\rm H_2}+M_{\rm H{\scriptscriptstyle  I}}$) budget of void galaxies together with the atomic gas mass (which is documented in the literature).

    In this paper we present the first  survey of $\rm H_2$ in void galaxies. It is a pilot survey for the CO Calar Alto Void Integral-field Treasury Survey (CO-CAVITY). CAVITY\footnote{https://cavity.caha.es/} is an integral-field unit (IFU) legacy survey for void galaxies. The CAVITY galaxies are currently  being observed with the PMAS-PPAK IFU of the Centro Astronómico Hispano en Andalucía (CAHA) to study the stellar populations, star formation histories, and stellar kinematics in void galaxies. A sub-sample of CAVITY galaxies will be observed in CO (CO-CAVITY). We present here the results of a pilot survey, consisting of the CO data of 20 void galaxies selected from the Void Galaxy Survey \citep[VGS, ][]{2012AJ....144...16K}, which will be contained in CAVITY and CO-CAVITY. We have observed the CO-CAVITY pilot survey with the 30 m antenna of the Institut de Radioastronomie Millimétrique (IRAM) at Pico Veleta to study the $\rm H_2$ and star formation properties of void galaxies. While the CAVITY and CO-CAVITY are getting started, this pilot research will give us first results of how star formation has developed in void galaxies.

    The present paper is organised into five sections and three appendices. In Section \ref{sec:sample} we present the selection of the CO-CAVITY pilot survey of galaxies in voids, and the control samples of galaxies in filaments and walls. In Section \ref{sec:results} we carry out a comparison of  different properties (such as sSFR, $M_{\rm H_2}$, SFE, $M_{\rm H{\scriptscriptstyle  I}}$, and molecular-to-atomic gas mass ratio) between these galaxy samples. In Section \ref{sec:discussion} we discus our results and compare them with previous studies. In Section \ref{sec:conclusions} we summarise our conclusions. In Appendix \ref{sec:theo-line-ratio} we estimate the theoretically expected CO line ratios. In Appendix \ref{sec:SFR} we compare different SFR tracers for galaxies in our samples. In Appendix \ref{sec:cospectra} we show the CO line spectra for the 20 observed galaxies. For this entire study, we assume a flat $\Lambda$CDM cosmology with a matter density $\rm \Omega=0.3$, a dark energy density $\Lambda=0.7$, and a Hubble constant $\rm H_0=70\,km\,s^{-1}\,Mpc^{-1}$.

\section{Sample and data}
    \label{sec:sample}

    \subsection{Void Galaxy Survey}
        The VGS is a volume-selected sample of 60 galaxies plus 18 companions that are located in the interior of the large-scale void regions of the cosmic web. It has been defined from the SDSS Data Release 7 (SDSS-DR7) in a volume from redshift $\rm z=0.003$ to $\rm z=0.030$. 
        The VGS is the first survey selected by a strictly geometric procedure based on the local spatial distribution of galaxies \citep{2011AJ....141....4K, 2011MNRAS.416.2494P}, where it is assumed that galaxies are good tracers of their surrounding density field. The relative density of a region in the Universe is calculated with the density contrast, $\delta=\rho/\rho_u-1$,  where $\rho$ is the density of the region, and $\rho_u$ is the mean density of the Universe. For the VGS, the density contrast ranges from -0.94 to -0.44 with a mean value of -0.78 \citep{2012AJ....144...16K}, and the size of the voids spans from $\rm 16.25 \,Mpc$ to $\rm 18.64\,Mpc$ \citep{2011AJ....141....4K}. These values of density contrast and size of the voids are comparable to other void galaxy surveys such as \cite{2012MNRAS.421..926P} with $\delta<-0.85$, and $\rm 17h^{-1}Mpc$ as the mean size of the voids.
   
        The VGS galaxy selection is based on the Delaunay tessellation field estimator \citep[DTFE, ][]{2000A&A...363L..29S, 2007PhDT.......433S, 2009LNP...665..291V}, which generates a continuous density field from the spatial distribution of the SDSS galaxies. This technique keeps the anisotropic structure of the cosmic web. The watershed void finder \citep[WVF, ][]{2007MNRAS.380..551P}, is applied to the density field to identify the void regions. The WVF is used in  geophysics to identify the basins of a landscape where the rainfall will collect. In the same way, it is applied to the DTFE density field to define the irregular and twisted voids boundaries. 
   
        The VGS galaxy selection criteria \citep{2011AJ....141....4K} are to be (i) located in the interior regions of voids and far from their boundaries, (ii) far from the SDSS volume limits, (iii) separated by at least $\rm 750\,km\,s^{-1}$ in velocity from a foreground and background cluster to avoid Finger of God effects, and (iv)  within a redshift of $\rm 0.010<z<0.025$ to select galaxies that are bright enough for H{\tiny I} observations \citep{2012AJ....144...16K}.
        
        The VGS has been defined without applying any colour or luminosity  selection. The only magnitude bias is the SDSS completeness limit at $r$-Petrosian < 17.77 mag \citep{2002AJ....124.1810S, 2015A&A...578A.110A}. This means that the sample is progressively less sensitive to faint objects with increasing redshift. The SDSS completeness limit corresponds to an absolute magnitude  of $M_{\rm r}=-17.4\,{\rm mag}$ at the maximum redshift of our sample ($z = 0.025$). For fainter objects, the sample is therefore not entirely complete. However, given the type of study that we carry out and given that the redshift range of the VGS sample is small,  we do not expect this to be a severe problem for this work.
  
    \subsubsection{CO sub-sample of the VGS \label{sec:cosubsamp}}
   
        We chose 20 galaxies for the CO observations (the CO-VGS sub-sample) from the galaxies in the VGS that were observed in H{\tiny I} by \cite{2012AJ....144...16K}. We based our selection on the SFR and $\rm M_\star$ from the Max-Planck-Institut für Astrophysik and Johns Hopkins University  \citep[MPA-JHU,][]{2003MNRAS.341...33K, 2004MNRAS.351.1151B, 2007ApJS..173..267S}. We dropped 15 objects (2 main VGS objects and 13 companions) with no data for the SFR or $\rm M_\star$ in MPA-JHU. 
        We estimated for each galaxy the expected molecular gas mass using the measured SFR assuming $\rm SFE = 10^{-9}\,yr^{-1}$ \citep{2011MNRAS.415...61S}.  From the predicted $M_{\rm H_2}$ , we then derived  the expected velocity-integrated $\rm CO(1-0)$ intensity  $I_{\rm CO(1-0)}$ with the IRAM 30 m telescope adopting a Galactic CO-to-$\rm H_2$ conversion factor, $\rm \alpha_{\rm CO}= 3.2 \, M_{\odot} (K\,km\,s^{-1} pc^{2})^{-1}$  \citep{2013ARA&A..51..207B}. Based on this estimation, we  selected the galaxies with an expected $I_{\rm CO(1-0)}\geqslant0.6\,{\rm Kkm\,s^{-1}}$, a limit below which the observations become prohibitively long. Thus, we excluded those objects that are expected to be beyond the detecting capacity of the IRAM 30 m telescope. In addition, we dropped one galaxy (VGS07) with relatively low stellar mass ($\rm 10^{7.7}\,M_{\odot}$) in order  to exclude faint and low-metallicity objects for which the detection of CO is difficult and the determination of the $M_{\rm H_2}$ is affected by the uncertainties in the $\alpha_{\rm CO}$ factor  \citep{2013ARA&A..51..207B}. 
   
\begin{figure}
            \centering
            \includegraphics[width=\hsize]{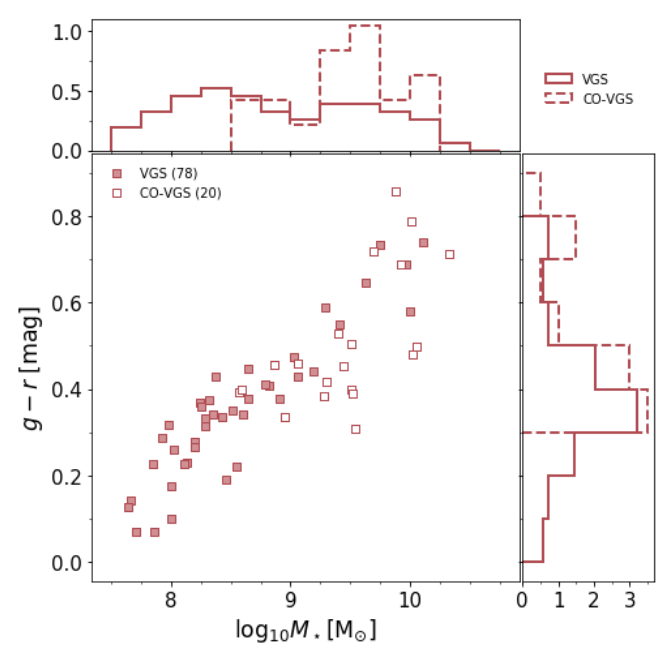}\\
            \caption{Colour vs. stellar mass diagram with normalised histograms for the VGS and the CO-VGS sub-sample.  The number of galaxies for each sample is shown in the legend.
            }
            \label{fig:col-mag}
       \end{figure}

        In Figure \ref{fig:col-mag} we show the colour-$M_\star$ diagram of the CO-VGS and the entire VGS. The CO-VGS $M_\star$ ranges from $\rm 10^{8.5}$ to $\rm 10^{10.3}\,M_{\odot}$, $g-r$ colour from 0.30 to 0.86 mag, SFR from $\rm 10^{-1.0}$ to $\rm 10^{0.7}\,M_{\odot}\,yr^{-1}$, and redshift from $\rm z=0.011$ to $\rm z=0.025$. The CO-VGS metallicity range is $8.44<\rm{12+\log(O/H)}< 9.10$ \citep[values taken from ][]{2004ApJ...613..898T}. The values of only four galaxies are lower than solar (8.66). Thus $\alpha_{\rm CO}$ was set to the Galactic conversion factor (without considering the  helium mass) of $\rm 3.2 \, M_{\odot} (K\,km\,s^{-1} pc^{2})^{-1}$ \citep{2013ARA&A..51..207B}.
   
        We need to be aware that our selection criterion  biases our sample against galaxies with a very low SFR (e.g. dwarf or elliptical galaxies). In addition, we might also miss galaxies with a low SFE that would have a higher molecular gas mass than we estimated and that might therefore be  erroneously excluded from our sample. 
   
    \subsubsection{Optical properties and atomic gas mass}
    \label{sec:opt-vgs}
   
        For the VGS we used photometric data from the SDSS-DR16, and spectrometric properties from the SDSS-DR16 \& MPA-JHU database \citep{2003MNRAS.341...33K, 2004MNRAS.351.1151B, 2004ApJ...613..898T, 2007ApJS..173..267S}. In particular, we used redshift (z), apparent dereddened magnitudes in $r$ and $g$ bands, $r$-Petrosian $R_{\rm 90}$, and $M_\star$. 
   
        For the SFR, we did not use the MPA-JHU values because there are systematic discrepancies between them and other SFR tracers, in particular for low SFRs ($\rm SFR \lesssim 10^{-1.0}\,M_\odot \, yr^{-1}$) and large radii ($R_{\rm 90} \gtrsim 15\, {\rm arcsec}$). A detailed analysis is given in Appendix \ref{sec:SFR}. 

        Instead, we used the SFR derived by \cite{2016MNRAS.458..394B} from $\rm H\alpha$ maps obtained at the 2.4 m Hiltner Telescope with the Echelle CCD in direct mode. The  $\rm H\alpha$ fluxes measured from the calibrated maps were extinction corrected based on the Balmer decrement derived from the central 3 arcsec spectra from the MPA-JHA DR7 catalogue and, in some cases, on infrared data from WISE and the 4000 $\AA$ break, $D_n(4000)$. The SFR was then calculated following \citet{2011ApJ...741..124H} and \citet{2011ApJ...737...67M} as 
    
        \begin{equation}
            \label{eq:sfr}
            SFR[M_\odot] = 5.4 \time 10^{-42} L_{\rm H\alpha} {\rm [erg\, s^{-1}]}
        .\end{equation}
   
        The SFR range of the VGS derived in this way is $\rm 10^{-2.7} \,M_\odot \, yr^{-1}<SFR<10^{0.8}\,M_\odot \, yr^{-1}$, and the range for the  CO-VGS sub-sample $\rm 10^{-2.1} M_\odot \, yr^{-1}<SFR<10^{0.8}\,M_\odot \, yr^{-1}$, which is slightly different from the range derived from the MPA-JHU SFR tracer (see Section \ref{sec:cosubsamp}). 
  
        Observations of the H{\tiny I} 21 cm line were obtained using the Westerbork Synthesis Radio Telescope (WSRT) for 73 of the 78 VGS galaxies. They were presented in \cite{2012AJ....144...16K}, and the reduction of these data was further explained in \cite{2011AJ....141....4K}. \cite{2012AJ....144...16K} derived the atomic gas mass ($M_{\rm H{\scriptscriptstyle  I}}$) using the luminosity distance calculated from the H{\tiny I} redshift. We re-scaled $M_{ \rm H{\scriptscriptstyle  I}}$ to the SDSS redshift luminosity distance that we use in the present paper.
   
        The data are listed in Tables \ref{tab:Opt} and \ref{tab:Spec} for the main galaxies in the VGS and in Tables \ref{tab:Opt_comp} and \ref{tab:Spec_comp} for the VGS companions.
   
        \begin{table}
            \small
            \caption{\label{tab:Opt}Photometric properties for the VGS.}
            \centering
            \begin{tabular}{|l|ccccc|}
                \hline
                \parbox[t]{1mm}{\multirow{3}{*}{\rotatebox[origin=c]{0}{ Name}}}
                &$g$&$r$&$g-r$&$r_{\rm 25}$&$i^{(*)}$\\
                &[mag]&[mag]&[mag]&[arcsec]&$[^\circ]$ \\
                &(1)&(2)&(3)&(4)&(5) \\
                \hline\hline
                VGS01 &    17.87 &    17.42 &  0.45 &   9.8 &  60.0 \\
                VGS02 &    17.72 &    17.34 &  0.38 &  11.8 &  47.0 \\
                VGS03 &    17.86 &    17.43 &  0.43 &   8.4 &  55.0 \\
                VGS04 &    17.22 &    16.76 &  0.46 &   7.0 &  46.0 \\
                VGS05 &    15.68 &    14.94 &  0.74 &  23.0 &  37.0 \\
                VGS06 &    17.67 &    17.34 &  0.33 &  12.5 &  71.0 \\
                VGS07 &    17.45 &    17.38 &  0.07 &  11.2 &  50.0 \\
                VGS08 &    24.57 &    22.87 &  1.71 &   1.5 &  51.0 \\
                VGS09 &    17.71 &    17.48 &  0.23 &  19.4 &  68.0 \\
                VGS10 &    17.20 &    16.86 &  0.34 &   1.8 &  87.0 \\
                VGS11 &    16.44 &    15.98 &  0.46 &  19.9 &  18.0 \\
                VGS12 &    17.64 &    17.36 &  0.28 &  11.1 &  29.0 \\
                VGS13 &    16.77 &    16.30 &  0.47 &  14.4 &  68.0 \\
                VGS14 &    17.00 &    16.73 &  0.26 &  18.8 &  67.0 \\
                VGS15 &    15.85 &    15.30 &  0.55 &  31.9 &     - \\
                VGS16 &    17.81 &    17.52 &  0.29 &  10.4 &  60.0 \\
                VGS17 &    20.97 &    21.09 & -0.12 &   1.8 &     - \\
                VGS18 &    17.82 &    17.44 &  0.38 &  17.6 &  82.0 \\
                VGS19 &    16.89 &    16.50 &  0.39 &   9.0 &  60.0 \\
                VGS20 &    18.15 &    18.01 &  0.14 &   5.2 &     - \\
                VGS21 &    15.49 &    14.81 &  0.69 &  43.8 &  79.0 \\
                VGS22 &    17.81 &    17.50 &  0.31 &   6.6 &  63.0 \\
                VGS23 &    15.53 &    15.14 &  0.38 &  22.4 &  49.0 \\
                VGS24 &    15.45 &    14.87 &  0.58 &  16.4 &  35.0 \\
                VGS25 &    17.83 &    17.60 &  0.23 &   9.7 &  43.0 \\
                VGS26 &    16.74 &    16.33 &  0.42 &  20.0 &  69.0 \\
                VGS27 &    18.04 &    17.72 &  0.32 &   9.1 &  55.0 \\
                VGS28 &    18.18 &    17.54 &  0.64 &  19.5 &  54.0 \\
                VGS29 &    16.60 &    16.23 &  0.38 &  15.6 &     - \\
                VGS30 &    18.13 &    17.96 &  0.17 &  18.8 &  77.0 \\
                VGS31 &    15.01 &    14.70 &  0.31 &  15.5 &  52.0 \\
                VGS32 &    14.57 &    14.12 &  0.45 &  27.0 &  46.0 \\
                VGS33 &    17.97 &    17.61 &  0.37 &   8.9 &  60.0 \\
                VGS34 &    16.11 &    15.26 &  0.86 &  16.7 &  50.0 \\
                VGS35 &    16.74 &    16.33 &  0.41 &  13.2 &  65.0 \\
                VGS36 &    16.74 &    16.41 &  0.33 &  18.2 &  79.0 \\
                VGS37 &    17.38 &    17.04 &  0.34 &  23.1 &  67.0 \\
                VGS38 &    17.05 &    16.98 &  0.07 &  16.9 &  39.0 \\
                VGS39 &    16.00 &    15.21 &  0.79 &  15.4 &  66.0 \\
                VGS40 &    17.25 &    16.82 &  0.43 &   9.0 &  42.0 \\
                VGS41 &    17.60 &    17.19 &  0.41 &   6.2 &  29.0 \\
                VGS42 &    16.33 &    15.80 &  0.53 &  14.8 &  58.0 \\
                VGS43 &    18.16 &    17.83 &  0.33 &   7.3 &  30.0 \\
                VGS44 &    15.20 &    14.80 &  0.40 &  17.1 &  31.0 \\
                VGS45 &    17.61 &    17.35 &  0.26 &  20.8 &  66.0 \\
                VGS46 &    17.13 &    16.78 &  0.35 &  14.1 &  71.0 \\
                VGS47 &    15.23 &    14.51 &  0.71 &  29.5 &  72.0 \\
                VGS48 &    17.61 &    17.02 &  0.59 &  11.4 &     - \\
                VGS49 &    15.85 &    15.46 &  0.39 &  12.3 &  40.0 \\
                VGS50 &    16.00 &    15.32 &  0.69 &  12.6 &  83.0 \\
                VGS51 &    17.25 &    17.03 &  0.22 &  12.8 &  63.0 \\
                VGS52 &    17.79 &    17.56 &  0.23 &  16.2 &  70.0 \\
                VGS53 &    16.12 &    15.61 &  0.50 &  20.5 &  64.0 \\
                VGS54 &    16.80 &    16.15 &  0.65 &  22.5 &  80.0 \\
                VGS55 &    16.63 &    16.19 &  0.44 &  18.1 &  55.0 \\
                VGS56 &    16.44 &    15.72 &  0.72 &  14.0 &  59.0 \\
                VGS57 &    14.94 &    14.44 &  0.50 &  21.5 &  30.0 \\
                VGS58 &    16.05 &    15.65 &  0.40 &  21.3 &  38.0 \\
                VGS59 &    18.12 &    17.76 &  0.36 &  12.7 &  67.0 \\
                VGS60 &    16.45 &    15.72 &  0.73 &  19.2 &  81.0 \\
                \hline                  
            \end{tabular}
            \tablefoot{Columns (1) and (2) are the dereddened $g$ and $r$-band magnitudes from SDSS. Column (3) is the dereddened $g-r$ colour. Column (4) is the radius of the galaxy at isophote 25, calculated using the $r$-Petrosian $R_{\rm 90}$ in the $r$ band from SDSS as \makebox{$r_{25}=1.7\times R_{90}$} (see Section \ref{sec:fap}). Column (5) is the galaxy inclination from \cite{2012AJ....144...16K}.
            
            (*) We only need the inclination for galaxies observed in CO to calculate the aperture correction factor.
            }
        \end{table}  

        \begin{table}
            \small
            \caption{\label{tab:Opt_comp}Photometric properties for the VGS companions.}
            \centering
            \begin{tabular}{|l|ccccc|}
                \hline
                \parbox[t]{1mm}{\multirow{3}{*}{\rotatebox[origin=c]{0}{ Name}}}
                &$g$&$r$&$g-r$&$r_{\rm 25}$&$i^{(*)}$\\
                &[mag]&[mag]&[mag]&[arcsec]&$[^\circ]$ \\
                &(1)&(2)&(3)&(4)&(5) \\
                \hline\hline
                VGS07a &    21.48 &    21.91 & -0.43 &   8.5 &     - \\
                VGS09a &    23.46 &    23.52 & -0.06 &   2.2 &     - \\
                VGS10a &    22.11 &    21.77 &  0.33 &   2.2 &     - \\
                VGS26a &    14.16 &    13.67 &  0.49 &  38.5 &     - \\
                VGS30a &    18.67 &    18.41 &  0.26 &  16.7 &     - \\
                VGS31a &    14.73 &    14.25 &  0.48 &  30.1 &  60.0 \\
                VGS31b &    16.93 &    16.74 &  0.19 &  10.8 &     - \\
                VGS34a &    20.38 &    20.28 &  0.10 &   5.6 &     - \\
                VGS36a &    19.88 &    19.88 & -0.00 &   2.5 &     - \\
                VGS37a &    16.57 &    16.22 &  0.35 &  16.9 &     - \\
                VGS38a &    17.69 &    17.57 &  0.13 &   7.8 &     - \\
                VGS38b &    19.03 &    18.86 &  0.17 &   9.8 &     - \\
                VGS39a &    19.39 &    19.44 & -0.05 &   6.7 &     - \\
                VGS51a &    22.28 &    20.70 &  1.58 &   5.5 &     - \\
                VGS53a &    17.61 &    16.65 &  0.96 &  19.1 &     - \\
                VGS54a &    19.32 &    18.98 &  0.34 &  11.2 &     - \\
                VGS56a &    19.19 &    18.90 &  0.29 &   7.9 &     - \\
                VGS57a &    17.85 &    17.75 &  0.10 &   7.7 &     - \\
                \hline                  
            \end{tabular}
            \tablefoot{Columns (1) and (2) are the dereddened $g$ and $r$-band magnitudes from SDSS derived in the same aperture. Column (3) is the dereddened $g-r$ colour magnitude from SDSS. Column (4) is the radius of the galaxy at isophote 25, calculated using the $r$-Petrosian $R_{\rm 90}$ in the $r$ band from SDSS as $r_{\rm 25}=1.7\times R_{\rm 90}$ (see Section \ref{sec:fap}). Column (5) is the galaxy inclination from HyperLEDA; there is no inclination data for VGS companions in  \cite{2012AJ....144...16K}. 

            (*) We only need the inclination for galaxies observed in CO to calculate the aperture correction factor.}
        \end{table}  
   
        \begin{table*}
            \small
            \caption{\label{tab:Spec}Spectrometric properties for the VGS.}
            \centering
            \begin{tabular}{|l|cccc|cccc|cccc|cc|}
                \hline
                \parbox[t]{1mm}{\multirow{3}{*}{\rotatebox[origin=c]{0}{ Name}}}
                &\multicolumn{4}{c|}{${\log_{10}M_\star}$ }&\multicolumn{4}{c|}{${\log_{10}M_{\rm H{\scriptscriptstyle  I}}}$ }&\multicolumn{4}{c|}{$\rm{\log_{10}SFR}$}&z&$D_{\rm L}$\\
                &  \multicolumn{4}{c|}{$[\rm{M_{\odot}}]$}& \multicolumn{4}{c|}{$[\rm{M_{\odot}}]$}  & \multicolumn{4}{c|}{$[\rm{M_{\odot}\,yr^{-1}}]$}&&$[\rm{Mpc}]$ \\
                &  \multicolumn{4}{c|}{(1)}& \multicolumn{4}{c|}{(2)}  & \multicolumn{4}{c|}{(3)}&(4)&(5) \\
                \hline\hline
                 VGS01 &           &        8.65 &      $\pm$ &            0.10 &       < &      8.32 &        &             - &          &     -1.28 &     $\pm$ &          0.11 &  0.019 &   80.5 \\
                VGS02 &           &        8.65 &      $\pm$ &            0.08 &         &      8.83 &    $\pm$ &          0.12 &          &     -1.57 &     $\pm$ &          0.14 &  0.023 &   98.3 \\
                VGS03 &           &        8.38 &      $\pm$ &            0.09 &       < &      8.27 &        &             - &          &     -1.96 &     $\pm$ &          0.12 &  0.017 &   73.2 \\
                VGS04 &           &        8.87 &      $\pm$ &            0.08 &       < &      8.23 &        &             - &          &     -1.13 &     $\pm$ &          0.04 &  0.016 &   70.4 \\
                VGS05 &           &       10.11 &      $\pm$ &            0.09 &       < &      8.51 &        &             - &          &     -1.38 &     $\pm$ &          0.29 &  0.022 &   97.7 \\
                VGS06 &           &        8.42 &      $\pm$ &            0.06 &         &      9.20 &    $\pm$ &          0.07 &          &     -1.17 &     $\pm$ &          0.10 &  0.023 &  100.4 \\
                VGS07 &           &        7.71 &      $\pm$ &            0.19 &         &      8.93 &    $\pm$ &          0.04 &          &     -1.10 &     $\pm$ &          0.03 &  0.016 &   70.5 \\
                VGS08 &           &           - &          &               - &         &      8.60 &    $\pm$ &          0.16 &          &     -1.52 &     $\pm$ &          0.13 &  0.020 &   85.3 \\
                VGS09 &           &        7.85 &      $\pm$ &            0.05 &         &      9.03 &    $\pm$ &          0.04 &          &     -1.43 &     $\pm$ &          0.04 &  0.013 &   56.2 \\
                VGS10 &           &        8.35 &      $\pm$ &            0.06 &         &      9.17 &    $\pm$ &          0.06 &          &     -1.59 &     $\pm$ &          0.02 &  0.016 &   68.4 \\
                VGS11 &           &        9.06 &      $\pm$ &            0.08 &         &      9.35 &    $\pm$ &          0.04 &          &     -2.05 &     $\pm$ &          0.24 &  0.016 &   71.5 \\
                VGS12 &           &        8.20 &      $\pm$ &            0.05 &         &      9.48 &    $\pm$ &          0.06 &          &     -1.35 &     $\pm$ &          0.13 &  0.018 &   77.3 \\
                VGS13 &           &        9.03 &      $\pm$ &            0.08 &         &      9.11 &    $\pm$ &          0.08 &          &     -1.14 &     $\pm$ &          0.14 &  0.019 &   83.1 \\
                VGS14 &           &        8.20 &      $\pm$ &            0.06 &         &      8.81 &    $\pm$ &          0.07 &          &     -1.74 &     $\pm$ &          0.07 &  0.013 &   56.9 \\
                VGS15 &           &        9.41 &      $\pm$ &            0.09 &         &         - &        &             - &          &     -0.70 &     $\pm$ &          0.32 &  0.019 &   82.9 \\
                VGS16 &           &        7.93 &      $\pm$ &            0.06 &       < &      8.06 &        &             - &          &     -2.10 &     $\pm$ &          0.05 &  0.013 &   57.6 \\
                VGS17 &           &           - &          &               - &         &         - &        &             - &          &     -1.52 &     $\pm$ &          0.09 &  0.011 &   46.2 \\
                VGS18 &           &        8.32 &      $\pm$ &            0.07 &         &      8.61 &    $\pm$ &          0.12 &          &     -2.70 &     $\pm$ &          0.00 &  0.016 &   71.0 \\
                VGS19 &           &        8.57 &      $\pm$ &            0.08 &         &      8.50 &    $\pm$ &          0.10 &          &     -1.25 &     $\pm$ &          0.09 &  0.014 &   62.6 \\
                VGS20 &           &        7.66 &      $\pm$ &            0.13 &         &         - &        &             - &          &     -1.15 &     $\pm$ &          0.06 &  0.017 &   72.1 \\
                VGS21 &           &        9.97 &      $\pm$ &            0.09 &         &      9.34 &    $\pm$ &          0.07 &          &     -0.64 &     $\pm$ &          0.34 &  0.017 &   75.3 \\
                VGS22 &           &        8.28 &      $\pm$ &            0.06 &       < &      8.42 &        &             - &          &     -1.66 &     $\pm$ &          0.06 &  0.019 &   83.3 \\
                VGS23 &           &        9.28 &      $\pm$ &            0.08 &         &      9.60 &    $\pm$ &          0.05 &          &     -0.98 &     $\pm$ &          0.12 &  0.017 &   72.4 \\
                VGS24 &           &       10.00 &      $\pm$ &            0.10 &       < &      8.53 &        &             - &          &     -0.08 &     $\pm$ &          0.43 &  0.023 &  101.2 \\
                VGS25 &           &        8.14 &      $\pm$ &            0.05 &         &      8.24 &    $\pm$ &          0.16 &          &     -0.95 &     $\pm$ &          0.08 &  0.019 &   82.6 \\
                VGS26 &           &        9.30 &      $\pm$ &            0.11 &         &      9.18 &    $\pm$ &          0.09 &          &     -1.10 &     $\pm$ &          0.16 &  0.023 &  101.1 \\
                VGS27 &           &        7.98 &      $\pm$ &            0.06 &         &      8.54 &    $\pm$ &          0.09 &          &     -2.22 &     $\pm$ &          0.00 &  0.015 &   64.7 \\
                VGS28 &           &           - &          &               - &       < &      8.22 &        &             - &          &         - &         &             - &  0.015 &   66.2 \\
                VGS29 &           &        8.91 &      $\pm$ &            0.07 &         &         - &        &             - &          &     -1.05 &     $\pm$ &          0.14 &  0.020 &   87.1 \\
                VGS30 &           &        8.00 &      $\pm$ &            0.07 &         &      8.77 &    $\pm$ &          0.07 &          &     -2.30 &     $\pm$ &          0.17 &  0.019 &   84.5 \\
                VGS31 &           &        9.55 &      $\pm$ &            0.09 &         &      9.31 &    $\pm$ &          0.06 &          &      0.31 &     $\pm$ &          0.00 &  0.021 &   91.0 \\
                VGS32 &           &        9.44 &      $\pm$ &            0.08 &         &      9.59 &    $\pm$ &          0.05 &          &     -0.64 &     $\pm$ &          0.08 &  0.012 &   51.1 \\
                VGS33 &           &        8.24 &      $\pm$ &            0.07 &         &      8.93 &    $\pm$ &          0.09 &          &         - &         &             - &  0.018 &   79.2 \\
                VGS34 &           &        9.88 &      $\pm$ &            0.10 &         &      9.39 &    $\pm$ &          0.05 &          &         - &         &             - &  0.017 &   71.7 \\
                VGS35 &           &        8.82 &      $\pm$ &            0.08 &         &      9.03 &    $\pm$ &          0.05 &          &     -1.07 &     $\pm$ &          0.16 &  0.017 &   75.1 \\
                VGS36 &           &        8.95 &      $\pm$ &            0.07 &         &      9.31 &    $\pm$ &          0.06 &          &     -0.80 &     $\pm$ &          0.14 &  0.022 &   97.6 \\
                VGS37 &           &        8.61 &      $\pm$ &            0.07 &         &      9.13 &    $\pm$ &          0.06 &          &     -1.36 &     $\pm$ &          0.11 &  0.019 &   84.1 \\
                VGS38 &           &        7.86 &      $\pm$ &            0.10 &         &      9.03 &    $\pm$ &          0.03 &          &     -1.30 &     $\pm$ &          0.00 &  0.014 &   59.8 \\
                VGS39 &           &       10.01 &      $\pm$ &            0.10 &       < &      8.41 &        &             - &          &     -0.85 &     $\pm$ &          0.16 &  0.019 &   82.7 \\
                VGS40 &           &        9.07 &      $\pm$ &            0.08 &         &      8.79 &    $\pm$ &          0.10 &          &     -0.73 &     $\pm$ &          0.10 &  0.024 &  103.4 \\
                VGS41 &           &        8.78 &      $\pm$ &            0.08 &       < &      8.46 &        &             - &          &     -1.09 &     $\pm$ &          0.12 &  0.023 &  101.9 \\
                VGS42 &           &        9.40 &      $\pm$ &            0.09 &         &      8.61 &    $\pm$ &          0.16 &          &     -0.82 &     $\pm$ &          0.15 &  0.019 &   81.5 \\
                VGS43 &           &        8.28 &      $\pm$ &            0.07 &       < &      8.46 &        &             - &          &     -1.59 &     $\pm$ &          0.13 &  0.021 &   93.4 \\
                VGS44 &           &        9.51 &      $\pm$ &            0.12 &         &      8.69 &    $\pm$ &          0.10 &          &     -0.21 &     $\pm$ &          0.10 &  0.018 &   76.6 \\
                VGS45 &           &        8.02 &      $\pm$ &            0.06 &         &      8.55 &    $\pm$ &          0.16 &          &     -2.22 &     $\pm$ &          0.00 &  0.015 &   63.0 \\
                VGS46 &           &        8.51 &      $\pm$ &            0.08 &         &      8.75 &    $\pm$ &          0.13 &          &     -1.35 &     $\pm$ &          0.04 &  0.016 &   69.0 \\
                VGS47 &           &       10.33 &      $\pm$ &            0.09 &         &      9.12 &    $\pm$ &          0.09 &          &     -0.10 &     $\pm$ &          0.18 &  0.022 &   96.5 \\
                VGS48 &           &        9.29 &      $\pm$ &            0.10 &         &         - &        &             - &          &     -1.12 &     $\pm$ &          0.19 &  0.025 &  109.0 \\
                VGS49 &           &        9.52 &      $\pm$ &            0.10 &       < &      8.56 &        &             - &          &     -0.22 &     $\pm$ &          0.09 &  0.025 &  108.8 \\
                VGS50 &           &        9.92 &      $\pm$ &            0.09 &         &      9.74 &    $\pm$ &          0.06 &          &     -0.66 &     $\pm$ &          0.16 &  0.020 &   88.6 \\
                VGS51 &           &        8.55 &      $\pm$ &            0.05 &         &      9.30 &    $\pm$ &          0.05 &          &     -0.31 &     $\pm$ &          0.04 &  0.025 &  110.5 \\
                VGS52 &           &        8.11 &      $\pm$ &            0.07 &         &      8.95 &    $\pm$ &          0.11 &          &     -1.70 &     $\pm$ &          0.20 &  0.018 &   78.2 \\
                VGS53 &           &        9.50 &      $\pm$ &            0.09 &         &      8.72 &    $\pm$ &          0.14 &          &     -0.64 &     $\pm$ &          0.14 &  0.021 &   93.5 \\
                VGS54 &           &        9.63 &      $\pm$ &            0.09 &         &      9.55 &    $\pm$ &          0.05 &          &     -0.76 &     $\pm$ &          0.12 &  0.024 &  104.2 \\
                VGS55 &           &        9.20 &      $\pm$ &            0.07 &         &      9.25 &    $\pm$ &          0.09 &          &     -0.83 &     $\pm$ &          0.09 &  0.025 &  109.8 \\
                VGS56 &           &        9.69 &      $\pm$ &            0.10 &       < &      8.43 &        &             - &          &     -0.63 &     $\pm$ &          0.16 &  0.019 &   81.3 \\
                VGS57 &           &       10.06 &      $\pm$ &            0.11 &         &      8.81 &    $\pm$ &          0.10 &          &      0.23 &     $\pm$ &          0.07 &  0.022 &   96.6 \\
                VGS58 &           &        8.59 &      $\pm$ &            0.06 &         &      8.87 &    $\pm$ &          0.04 &          &     -1.34 &     $\pm$ &          0.08 &  0.012 &   49.8 \\
                VGS59 &           &        8.26 &      $\pm$ &            0.07 &       < &      8.41 &        &             - &          &     -1.85 &     $\pm$ &          0.06 &  0.019 &   82.7 \\
                VGS60 &           &        9.75 &      $\pm$ &            0.09 &         &      8.41 &    $\pm$ &          0.27 &          &     -0.94 &     $\pm$ &          0.12 &  0.020 &   85.4 \\
                \hline                  
            \end{tabular}
            \tablefoot{Column (1) is the stellar mass from MPA-JHU \citep{2003MNRAS.341...33K, 2007ApJS..173..267S}. Column (2) is the atomic gas mass from WSRT \citep{2012AJ....144...16K}; it has been re-scaled to the luminosity distance in Column (5). Column (3) is the $\rm H\alpha$ based SFR from \cite{2016MNRAS.458..394B}. Column (4) is the redshift from MPA-JHU. Column (5) is the luminosity distance.}
        \end{table*}  

        \begin{table*}
            \small
            \caption{\label{tab:Spec_comp} Spectrometric properties for the VGS companions.}
            \centering
            \begin{tabular}{|l|cccc|cccc|cccc|cc|}
                \hline
                \parbox[t]{1mm}{\multirow{3}{*}{\rotatebox[origin=c]{0}{ Name}}}
                &\multicolumn{4}{c|}{${\log_{10}M_\star}$ }&\multicolumn{4}{c|}{${\log_{10}M_{\rm H{\scriptscriptstyle  I}}}$ }&\multicolumn{4}{c|}{$\rm{\log_{10}SFR}$}&z&$D_{\rm L}$\\
                &\multicolumn{4}{c|}{$[\rm{M_{\odot}}]$}& \multicolumn{4}{c|}{$[\rm{M_{\odot}}]$}& \multicolumn{4}{c|}{$[\rm{M_{\odot}\,yr^{-1}}]$}&&$[\rm{Mpc}]$ \\
                &\multicolumn{4}{c|}{(1)}& \multicolumn{4}{c|}{(2)}  & \multicolumn{4}{c|}{(3)}&(4)&(5) \\
                \hline\hline
                VGS07a &           &           - &          &               - &         &      8.52 &    $\pm$ &          0.09 &          &         - &         &             - &  0.016 &   70.5 \\
                VGS09a &           &           - &          &               - &         &      7.79 &    $\pm$ &          0.10 &          &         - &         &             - &  0.013 &   56.2 \\
                VGS10a &           &           - &          &               - &         &      8.93 &    $\pm$ &          0.10 &          &         - &         &             - &  0.016 &   68.4 \\
                VGS26a &           &           - &          &               - &         &     10.31 &    $\pm$ &          0.04 &          &         - &         &             - &  0.023 &  101.1 \\
                VGS30a &           &           - &          &               - &         &      8.71 &    $\pm$ &          0.07 &          &         - &         &             - &  0.019 &   84.5 \\
                VGS31a &           &       10.02 &      $\pm$ &            0.14 &         &      9.26 &    $\pm$ &          0.05 &          &      0.81 &     $\pm$ &          0.00 &  0.021 &   91.0 \\
                VGS31b &           &        8.47 &      $\pm$ &            0.05 &         &      8.45 &    $\pm$ &          0.15 &          &     -0.82 &     $\pm$ &          0.00 &  0.021 &   91.0 \\
                VGS34a &           &           - &          &               - &         &      7.71 &    $\pm$ &          0.13 &          &         - &         &             - &  0.017 &   71.7 \\
                VGS36a &           &           - &          &               - &       < &     11.83 &        &             - &          &         - &         &             - &  0.022 &   97.6 \\
                VGS37a &           &           - &          &               - &         &      9.19 &    $\pm$ &          0.05 &          &     -1.15 &     $\pm$ &          0.07 &  0.019 &   84.1 \\
                VGS38a &           &        7.65 &      $\pm$ &            0.05 &         &      8.02 &    $\pm$ &          0.06 &          &     -1.74 &     $\pm$ &          0.07 &  0.014 &   59.8 \\
                VGS38b &           &           - &          &               - &         &      8.21 &    $\pm$ &          0.06 &          &     -2.30 &     $\pm$ &          0.00 &  0.014 &   59.8 \\
                VGS39a &           &           - &          &               - &         &      8.50 &    $\pm$ &          0.05 &          &         - &         &             - &  0.019 &   82.7 \\
                VGS51a &           &           - &          &               - &         &      8.37 &    $\pm$ &          0.04 &          &         - &         &             - &  0.025 &  110.5 \\
                VGS53a &           &           - &          &               - &         &      8.58 &    $\pm$ &          0.25 &          &         - &         &             - &  0.021 &   93.5 \\
                VGS54a &           &           - &          &               - &       < &     11.65 &        &             - &          &         - &         &             - &  0.024 &  104.2 \\
                VGS56a &           &           - &          &               - &         &      8.25 &    $\pm$ &          0.16 &          &         - &         &             - &  0.019 &   81.3 \\
                VGS57a &           &        8.00 &      $\pm$ &            0.07 &         &      8.41 &    $\pm$ &          0.08 &          &         - &         &             - &  0.022 &   96.6 \\
                \hline                  
            \end{tabular}
            \tablefoot{Column (1) is the stellar mass from MPA-JHU \citep{2003MNRAS.341...33K, 2007ApJS..173..267S}. Column (2) is the atomic gas mass from WSRT \citep{2012AJ....144...16K}; it has been re-scaled to the luminosity distance in column (5). Column (3) is the $\rm H\alpha$ based SFR from \cite{2016MNRAS.458..394B}. Column (4) is the redshift from MPA-JHU. Column (5) is the luminosity distance.}
        \end{table*}  
   
   \subsubsection{CO observations and data reduction}
         
        The observations were carried out at the IRAM 30 m telescope in three different periods ($18^{\rm }$ - $23^{\rm }$ June, $11^{\rm }$ - $22^{\rm }$ July, and $17^{\rm }$ - $18^{\rm }$ October 2019). We observed the $\rm{^{12}CO(1-0)}$ (rest frame frequency 115.2712 GHz) emission line in parallel with the $\rm{^{12}CO(2-1)}$ (rest frame frequency 230.5380 GHz) emission line. 
         
        The EMIR dual-polarisation receiver was combined with two autocorrelators: FTS (with a  frequency resolution of 0.195 MHz, corresponding to a velocity resolution of $\rm 0.5\,km\,s^{-1}$ at \makebox{113 GHz)}, and WILMA (with frequency and velocity resolutions  of 2 MHz and $\rm 5.3\,km\,s^{-1}$). 
        We used the wobbler-switching method for the sky subtraction with a wobbler throw of 60-80 arcsec. This was chosen for each individual galaxy, checking their optical images (SDSS $g$-band) to ensure that the off-position was empty of emission.
         
        The bandwidths of the receiver (EMIR 16 GHz) and the backends (FTS 8 GHz, and WILMA 4 GHz) are wide enough to encompass the redshifted CO lines within one centrally tuned frequency setup (even though in the case of WILMA, some of the CO(2-1) lines lie very close to the edge of the bandwidth). The CO-VGS redshift ranges from $\rm z=0.011$ to $\rm z=0.025$, the recession velocities (optical convention) from 3454 $\rm km\,s^{-1}$ to 7446 $\rm km\,s^{-1}$, and the redshifted frequencies range from 112 to 114 GHz for $\rm CO(1-0)$ and from 225 to 228 GHz for $\rm CO(2-1)$. According to this, the backends were tuned to an intermediate recession velocity of 5200 $\rm km\,s^{-1}$, which translates into redshifted frequencies of 113.3059 GHz for $\rm CO(1-0)$ and 226.6074 GHz for $\rm CO(2-1)$. We used the FTS spectra in this study because of their broader bandwidth and took the WILMA spectra only as a backup.
         
        Weather conditions were generally good for all the observations, except for $22^{\rm }$  June, when the pointing discrepancy was up to 10 arcsec. After removing this data set, the mean system temperature was 185 K in terms of ${T^*_{\rm A}}$ (antenna temperature with atmospheric correction) for $\rm CO(1-0)$, and \makebox{528 K} for $\rm CO(2-1)$. The pointing was checked every $\sim$ 1.5 hours by observing a close-by quasar, and its accuracy was better than \makebox{3-6 arcsec}. This is acceptable for the $\rm CO(1-0)$ with a half-power beam size of 22 arcsec at our observing frequency, but is not ideal for $\rm CO(2-1)$ with a half-power beam size of 11 arcsec. The planet Mars was observed every 2-3 hours to calibrate the antenna focus.
         
        The on-source observing time ranged from 30 minutes for the brightest sources to 2 hours for the faintest sources. The observations were generally carried out until the $\rm CO(1-0)$ line was detected with a signal-to-noise ratio ($\rm S/N$) greater than 5, except for VGS42, for which an S/N of only 3.8 could be achieved. If not detected, the observations were stopped at a root-mean-square noise ($rms$) of main beam temperature (${T_{\rm mb}}$) below 1.5 mK at a velocity resolution of 20 $\rm km\,s^{-1}$. 
         
        The line temperature is expressed in terms of ${T_{\rm mb}=T^*_{\rm A} \times (F_{\rm eff}/B_{\rm eff})}$, where ${F_{\rm eff}}$ is the IRAM 30 m telescope forward efficiency, which is 0.95 for $\rm CO(1-0)$ and 0.91 for $\rm CO(2-1)$, and ${B_{\rm eff}}$ is the beam efficiency, which is 0.77 for $\rm CO(1-0)$ and 0.58 for $\rm CO(2-1)$.
         
        We used the GILDAS\footnote{http://www.iram.fr/IRAMFR/GILDAS} software, provided by IRAM, to reduce the CO data. We discarded bad scans from the observations. We then subtracted a linear baseline from every spectrum. In some cases, spectra from the FTS backend have platforming, that is, the baseline of the spectrum presents steps at the end of the correlator bands. We used the  {\tt FtsPlatformingCorrection5.class}  program, developed by IRAM, to correct for this artefact. We then averaged the spectra and smoothed them to a spectral resolution of 20 $\rm km\,s^{-1}$. The final spectra are presented in Figs. \ref{fig:spec10} and \ref{fig:spec21}. 
        
        We derived the emission line intensity (${I_{\rm CO}}$) as the velocity-integrated ${T_{\rm mb}}$ within the zero-level line width (${\Delta V}$), which was determined visually for each averaged spectrum, 
        
        \begin{equation}\label{eq:icotmb}
            {I_{\rm CO}= \int_{\Delta V}^{}T_{\rm mb}(V){\rm d}V}\quad.
        \end{equation}
        
        For non-detections, upper limits were defined as \makebox{$I_{\rm CO}<3\times rms \times\sqrt{\delta V\Delta V}$}, where $\delta V$ is the channel width in $\rm km\,s^{-1}$, and $\Delta V$ was set to the mean value of the detections,  which is 300 $\rm km\,s^{-1}$ for $\rm CO(1-0)$ and 240 $\rm km\,s^{-1}$ for $\rm CO(2-1)$.  The 20 observed CO intensities and their statistical errors, calculated as $rms \times\sqrt{\delta V\Delta V}$, are listed in Table \ref{tab:ICO}. In addition to the statistical error, we took a typical calibration error of 15\% for $\rm CO(1-0)$ and 30\% for $\rm CO(2-1)$ into account, estimated by  \citet{2019A&A...627A.107L} from a comparison of the observation of four strong sources on different days. The $\rm CO(1-0)$ line was detected ($\rm S/N>3$) for 13 galaxies and the $\rm CO(2-1)$ line for 14 galaxies.
         
        \begin{table*}
            \caption{\label{tab:ICO}CO emission line intensities for the CO-VGS.}
            \centering
            \begin{tabular}{|l|ccccccc|ccccccc|}     % 7 columns 
                \hline       
                \parbox[t]{1mm}{\multirow{3}{*}{\rotatebox[origin=c]{0}{ Name}}}
                &\multicolumn{4}{c}{$I_{\rm CO(1-0)}$} & $rms$  & $S/N$ & $\Delta V_{\rm CO(1-0)}$ &   \multicolumn{4}{c}{$I_{\rm CO(2-1)}$}&$rms$ & $S/N$ & $\Delta V_{\rm CO(2-1)}$ \\
                &\multicolumn{4}{c}{[$\rm K\,km\,s^{-1}$]} &  [mK] & & [$\rm km\,s^{-1}$]&     \multicolumn{4}{c}{[$\rm K\,km\,s^{-1}$]} &[mK]& &[$\rm km\,s^{-1}$]\\
                &\multicolumn{4}{c}{(1)}  & (2)  & (3)  &  (4)& \multicolumn{4}{c}{(5)}&(6) & (7) & (8) \\
                \hline \hline
                VGS04 &      < &   0.34 &       &          - &    1.4 &       - &           300 &      < &   0.82 &       &          - &    3.9 &       - &           240 \\
                VGS11 &      < &   0.31 &       &          - &    1.3 &       - &           300 &        &   1.15 &   $\pm$ &       0.26 &    2.8 &     4.4 &           440 \\
                VGS19 &      < &   0.35 &       &          - &    1.5 &       - &           300 &      < &   0.68 &       &          - &    3.2 &       - &           240 \\
                VGS23 &        &   0.60 &   $\pm$ &       0.10 &    1.3 &     6.0 &           290 &        &   1.36 &   $\pm$ &       0.20 &    2.5 &     7.0 &           300 \\
                VGS26 &      < &   0.25 &       &          - &    1.0 &       - &           300 &      < &   0.58 &       &          - &    2.8 &       - &           240 \\
                VGS31 &        &   0.57 &   $\pm$ &       0.10 &    1.4 &     5.9 &           250 &        &   1.39 &   $\pm$ &       0.20 &    2.7 &     6.9 &           260 \\
                VGS31a &        &   2.65 &   $\pm$ &       0.20 &    2.3 &    13.6 &           350 &        &   2.15 &   $\pm$ &       0.34 &    5.2 &     6.3 &           210 \\
                VGS32 &        &   2.20 &   $\pm$ &       0.16 &    2.5 &    13.8 &           200 &        &   1.59 &   $\pm$ &       0.28 &    4.4 &     5.6 &           200 \\
                VGS34 &        &   3.22 &   $\pm$ &       0.22 &    2.6 &    14.3 &           360 &        &   3.38 &   $\pm$ &       0.42 &    4.9 &     8.1 &           350 \\
                VGS36 &      < &   0.32 &       &          - &    1.3 &       - &           300 &      < &   0.54 &       &          - &    2.6 &       - &           240 \\
                VGS39 &        &   0.60 &   $\pm$ &       0.11 &    1.2 &     5.4 &           380 &        &   0.89 &   $\pm$ &       0.24 &    2.5 &     3.8 &           420 \\
                VGS42 &        &   0.58 &   $\pm$ &       0.16 &    2.0 &     3.8 &           310 &      < &   0.70 &       &          - &    3.3 &       - &           240 \\
                VGS44 &        &   0.55 &   $\pm$ &       0.09 &    1.6 &     6.3 &           150 &        &   1.48 &   $\pm$ &       0.17 &    2.6 &     8.8 &           200 \\
                VGS47 &        &   2.28 &   $\pm$ &       0.17 &    1.8 &    13.8 &           420 &        &   1.30 &   $\pm$ &       0.25 &    3.3 &     5.2 &           280 \\
                VGS49 &        &   0.87 &   $\pm$ &       0.13 &    1.6 &     6.5 &           350 &        &   0.64 &   $\pm$ &       0.19 &    3.3 &     3.4 &           160 \\
                VGS50 &      < &   0.35 &       &          - &    1.5 &       - &           300 &        &   1.21 &   $\pm$ &       0.24 &    2.6 &     5.0 &           420 \\
                VGS53 &        &   0.60 &   $\pm$ &       0.09 &    1.2 &     6.6 &           300 &        &   0.72 &   $\pm$ &       0.21 &    3.2 &     3.4 &           200 \\
                VGS56 &        &   0.79 &   $\pm$ &       0.11 &    1.3 &     7.1 &           350 &        &   1.25 &   $\pm$ &       0.21 &    2.6 &     6.1 &           300 \\
                VGS57 &        &   3.19 &   $\pm$ &       0.16 &    2.0 &    20.5 &           300 &        &   4.38 &   $\pm$ &       0.23 &    3.5 &    19.2 &           200 \\
                VGS58 &      < &   0.29 &       &          - &    1.2 &       - &           300 &      < &   0.42 &       &          - &    2.0 &       - &           240 \\
                \hline                  
            \end{tabular}
            \tablefoot{(1) Velocity-integrated intensity and its statistical error of the $\rm CO(1-0)$ emission line. (2) Root-mean-square noise of the $\rm CO(1-0)$ emission line spectrum at a velocity resolution of $\rm 20\,km\,s^{-1}$. (3) Signal-to-noise ratio of the $\rm CO(1-0)$ emission line. (4) Spectral zero-level line width of the $\rm CO(1-0)$ emission line. For non-detections, we set it to the mean value of detections (300 $\rm km\,s^{-1}$). (5)-(8) The same for the $\rm CO(2-1)$ emission line. For non-detections in $\rm CO(2-1),$ we set the spectral zero-level line width of the emission line (8) to the mean value of detections (240 $\rm km\,s^{-1}$).}
        \end{table*}

    \subsubsection{Aperture correction} 
        \label{sec:fap}
          
        The IRAM 30 m telescope beam of $\rm CO(1-0)$ (22 arcsec) covers the entire galaxy in most objects of our sample. In order to correct for a small fraction of missing flux, we calculated an aperture correction following the procedure of \cite{2011A&A...534A.102L}. The resulting aperture correction factor ($f_{\rm ap}$), listed in Table \ref{tab:MH2}, has values between 1.1 and 1.5, and its mean value is 1.3. 
         
        The method assumes a molecular disc following an exponential profile with an exponential scale length $r_{\rm e}=0.2\times r_{\rm 25}$. Since $r_{\rm 25}$ is not available in HyperLEDA\footnote{http://leda.univ-lyon1.fr} \citep{1991A&A...243..319P} for all the objects and some values looked erroneous after visual inspection, we used the $r$-Petrosian $R_{\rm 90}$ from SDSS. We compared $R_{90}$ and $r_{\rm 25}$ for the objects for whichd both values exist, and find the ratio $r_{\rm 25}/ R_{\rm 90} = (1.7\pm0.5)$. We thus use $ R_{\rm 90}$ and the relation $r_{\rm 25} = 1.7\times R_{\rm 90}$. In order to calculate $f_{\rm ap}$, we furthermore need to know the inclination ($i$) of the galaxy, which we took from \cite{2012AJ....144...16K}, who performed a photometric analysis of the VGS. The values are listed in Tab.~\ref{tab:Opt}. An inclination is available in the literature for all galaxies in the CO-VGS sample.
          
    \subsubsection{Molecular gas mass }
        \label{sec:molmass}
         
        With the (main beam) temperature-to-flux conversion factor ($K_{\rm i-s}=5\,Jy\,K^{-1}$) of the IRAM 30 m telescope, the CO velocity integrated flux density is \makebox{$S_{\rm CO}\Delta V\,[{\rm Jy\,km\,s^{-1}}]=K_{\rm i-s} \times I_{\rm CO}\,[{\rm K\,km\,s^{-1}}]$.}
         
        Following \cite{1997ApJ...478..144S}, we calculated the \makebox{$\rm CO(1-0)$} luminosity as        
        $$L'_{\rm CO}[{\rm K\,km\,s^{-1}\,pc^{2}}]=3.25 \times 10^7 S_{\rm CO}\Delta V\nu_{\rm rest}^{-2}D_{\rm L}^2(1+{\rm z})^{-1}\quad, $$
        where $\nu_{\rm rest}$ is the emission line rest frequency in  GHz, $D_{\rm L}$ is the luminosity distance in Mpc, and $\rm z$ is the optical redshift from MPA-JHU.
         
        Finally, we calculated the molecular gas mass as \makebox{$M_{\rm H_2}= \alpha_{\rm CO}L_{\rm CO}$}, where $\alpha_{\rm CO}$ is the CO-to-$\rm H_2$ conversion factor. We considered the Galactic $\alpha_{\rm CO}= 3.2\,{\rm M_{\odot}\,(K\,km\,s^{-1}\,pc^{2})^{-1}}$ \citep{2013ARA&A..51..207B}, without correction for helium, as a constant value for the CO-VGS galaxies. Two galaxies (VGS11 and VGS50) are undetected in $\rm CO(1-0)$, but detected in $\rm CO(2-1)$. This is not unusual for objects that are smaller than the CO(1-0) beam for which the beam dilution is less severe for $I_{\rm CO(2-1)}$ than for $I_{\rm CO(1-0)}$ because the CO(2-1) beam size is smaller (see Sect.~\ref{sec:line-ratio} and Appendix \ref{sec:theo-line-ratio} for a detailed discussion of the influence of the beam size).
        For these cases, we estimated the $\rm CO(1-0)$ velocity-integrated intensity using the theoretical value of $R_{\rm 21theo} =  I_{\rm CO(2-1)}/I_{\rm CO(1-0)}$, calculated in Appendix \ref{sec:theo-line-ratio}. We adopted an intrinsic brightness ratio of $T_{\rm Bc2-1}/{T_{\rm Bc1-0}} = 0.8$ \citep{2009AJ....137.4670L}, and based on the corresponding values of $r_{\rm e}$ and $i$ for each galaxy, we derived $R_{\rm 21theo} = 3.0$ and $2.8$ for VGS11 and VGS50, respectively. The resulting molecular gas masses are listed in Table \ref{tab:MH2}.
   
        \begin{table}
            \caption{\label{tab:MH2}Molecular gas mass.}
            \centering
            \begin{tabular}{|l|ccccc|}
                \hline
                &  \multicolumn{4}{c}{$\log_{10}M_{\rm H_2}$}&       $f_{\rm ap}$ \\
                Name  &  \multicolumn{4}{c}{$[\rm{M_{\odot}}]$}&         \\
                &  \multicolumn{4}{c}{(1)}&(2) \\
                \hline\hline
                VGS04 &      < &      7.83 &       &             - & 1.05 \\
                VGS11 &        &      8.13 &   $\pm$ &          0.16 & 1.43 \\
                VGS19 &      < &      7.76 &       &             - & 1.07 \\
                VGS23 &        &      8.23 &   $\pm$ &          0.10 & 1.40 \\
                VGS26 &      < &      8.08 &       &             - & 1.25 \\
                VGS31 &        &      8.34 &   $\pm$ &          0.10 & 1.20 \\
                VGS31a &        &      9.12 &   $\pm$ &          0.07 & 1.57 \\
                VGS32 &        &      8.54 &   $\pm$ &          0.07 & 1.57 \\
                VGS34 &        &      8.90 &   $\pm$ &          0.07 & 1.23 \\
                VGS36 &      < &      8.14 &       &             - & 1.19 \\
                VGS39 &        &      8.26 &   $\pm$ &          0.10 & 1.16 \\
                VGS42 &        &      8.24 &   $\pm$ &          0.13 & 1.17 \\
                VGS44 &        &      8.21 &   $\pm$ &          0.09 & 1.30 \\
                VGS47 &        &      9.07 &   $\pm$ &          0.07 & 1.46 \\
                VGS49 &        &      8.66 &   $\pm$ &          0.09 & 1.15 \\
                VGS50 &        &      8.22 &   $\pm$ &          0.16 & 1.10 \\
                VGS53 &        &      8.41 &   $\pm$ &          0.09 & 1.28 \\
                VGS56 &        &      8.36 &   $\pm$ &          0.09 & 1.15 \\
                VGS57 &        &      9.22 &   $\pm$ &          0.07 & 1.45 \\
                VGS58 &      < &      7.60 &       &             - & 1.41 \\
                \hline                  
            \end{tabular}
            \tablefoot{Column (1) is the molecular gas mass and total error. Column (2) is the aperture correction factor. 

            (*) The molecular gas mass of these galaxies has been derived from the $\rm CO(2-1)$ line emission intensity and the theoretical $\rm CO(2-1)$-to-$\rm CO(1-0)$ line ratio estimated in Appendix \ref{sec:theo-line-ratio}.
            }
        \end{table}  
  
    \subsection{Control sample 
        \label{sec:contsamp}}
        For this study, we needed a comparison sample with $M_{\rm H_2}$, $M_{\rm H{\scriptscriptstyle  I}}$, $M_\star$, and SFR data. The comparison sample used in \cite{2012AJ....144...16K} cannot be used here because it does not contain any $\rm H_2$ data. We combined two $\rm H_2$ surveys: xCOLD GASS \citep{2017ApJS..233...22S} \citep[together with xGASS,][ for the H{\tiny I} data]{2018MNRAS.476..875C}, and EDGE-CALIFA \citep{2017ApJ...846..159B} (together with López-Sánchez et al. in prep. for the H{\tiny I} data) as comparison samples. These surveys do not have any environmental selection criteria, so they contain galaxies in voids, filaments, walls, and cluster. We wished to compare the CO-VGS galaxies with galaxies in filaments and walls alone, so we removed the galaxies from the control sample that inhabit voids and clusters (more details below). 
        For the optical properties, we used data from the MPA-JHU catalogue in the same way as for the VGS galaxies (see Section \ref{sec:opt-vgs}). In particular, we used the $M_\star$ \citep{2003MNRAS.341...33K, 2007ApJS..173..267S}, metallicity \citep{2004ApJ...613..898T}, and apparent dereddened magnitudes in $r$ and $g$ bands. We explain the choice of the SFR tracer in Appendix \ref{sec:SFR}.
   
    \subsubsection{Selection overview}
    \label{sec:selection-overview}
   
        From the 1690 galaxies in the xCOLD GASS  and López-Sánchez et al. (in prep.) samples, we  first removed galaxies inhabiting voids and clusters  by excluding objects  listed in the \cite{2012MNRAS.421..926P} void galaxy survey  and the \cite{2017A&A...602A.100T} group of galaxies survey (considering galaxies in groups with more than 30 members as a cluster).
   
\begin{figure}
            \centering
            \includegraphics[width=\hsize]{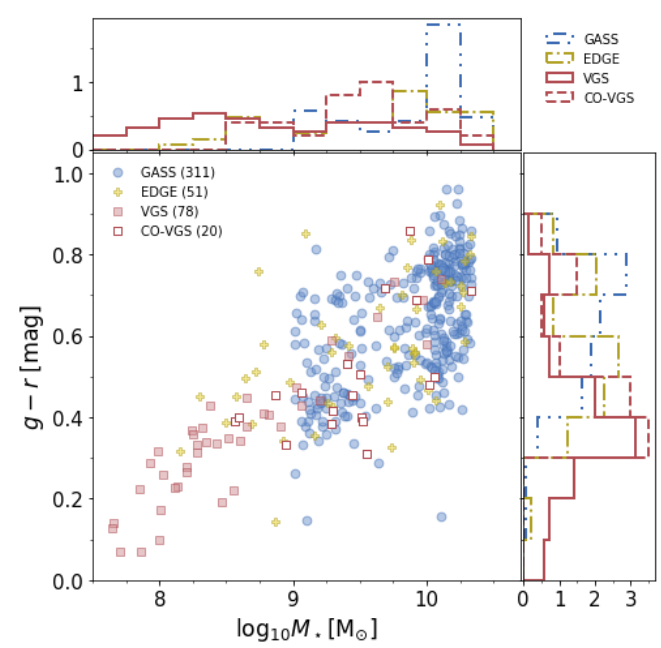}\\
            \caption{Colour vs. stellar mass diagram with normalised histograms for the VGS and the CCS. The number of galaxies for each sample is shown in the legend.}
            \label{fig:comp-col-mag}
        \end{figure}

\begin{figure}
            \centering
            \includegraphics[width=\hsize]{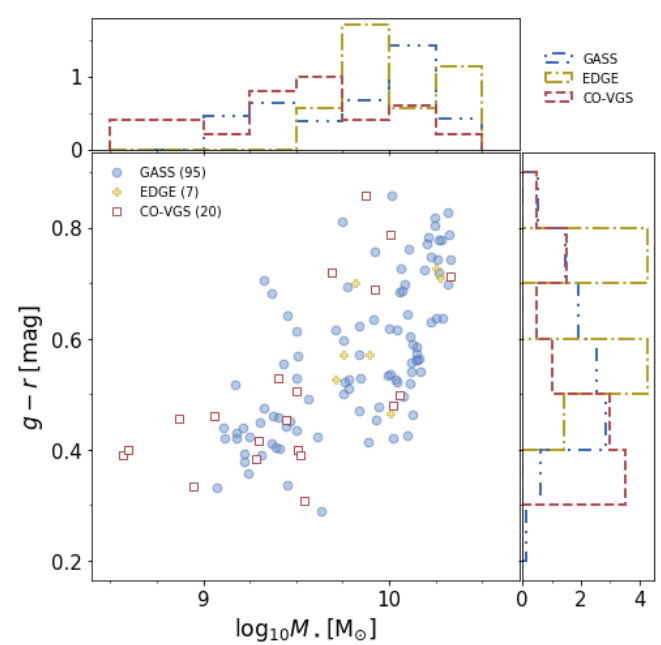}\\
            \caption{Colour vs. stellar mass diagram with normalised histograms for the CO-VGS and the CO-CS. The number of galaxies for each sample is shown in the legend.}
            \label{fig:trunc-col-mag}
        \end{figure}

        We then generated two sub-samples: one to compare with the entire VGS, and the other to compare with the CO-VGS. For the first sub-sample, called the complete control sample (CCS), we selected 362  galaxies that lie within the $M_\star$ and $g-r$ colour ranges of the entire VGS ($ 10^{7.7}\,{\rm M_{\odot}}<M_\star<10^{10.3}\,{\rm M_{\odot}}$ and $\rm 0.07\,mag<g-r<0.86\,mag$). For the second sub-sample, which we call CO comparison sample (CO-CS), we selected 102 galaxies with molecular gas data that lie within the $M_\star$, $g-r$ colour, and SFR ranges of the CO-VGS ($ 10^{8.5}\,{\rm M_{\odot}}<M_\star<10^{10.3}\,{\rm M_{\odot}}$, $0.30\,{\rm mag}<g-r<0.86\,{\rm mag}$, and $\rm SFR>0.1\,M_{\odot}\,yr^{-1}$). Figures \ref{fig:comp-col-mag} and \ref{fig:trunc-col-mag} show the colour-stellar mass distribution of the VGS and the CO-VGS, respectively. The CCS and CO-CS do not cover the $M_\star$ range below $\rm 10^{8.0}\,M_{\odot}$ and $\rm 10^{9.0}\,M_{\odot}$, respectively, thus the statistical comparison is only representative above these values. There are other $\rm H_2$ samples with low stellar mass galaxies \citep{2014A&A...563A..31R, 2020A&A...643A.180H, 2021arXiv210104389C}, but they were not useful for our study because there are only very few galaxies with ${\rm 10^{8.5}\,M_{\odot}}<M_\star<{\rm 10^{9.0}\,M_{\odot}}$, and the $M_{\rm H_2}$ values are highly dispersed. It is difficult to obtain $M_{\rm H_2}$ for galaxies with $M_\star<{\rm 10^{9.0}\,M_{\odot}}$ because their metallicities are low, which translates into low CO emission and a high uncertainty in the $\alpha_{\rm CO}$ value.
   
    \subsubsection{xGASS and xCOLD GASS surveys}
   
        The xGASS \citep[][the extended GALEX Arecibo SDSS Survey]{2018MNRAS.476..875C} is a H{\tiny I} survey of 1179 galaxies selected by $M_\star$ and redshift (${\rm 10^{9.0}\,M_{\odot}}<M_{\star}<{\rm 10^{11.5}\,M_{\odot}}$, and \makebox{$\rm 0.01<z<0.05$)}.
        
        xCOLD GASS \citep{2017ApJS..233...22S} is an IRAM 30 m telescope $\rm H_2$ legacy survey of 532 nearby stellar mass selected galaxies ($\rm 0.01<z<0.05$, and ${\rm 10^{9.0}\,M_{\odot}}<M_{\star}{\rm <10^{11.5}\,M_{\odot}}$). The metallicity of the galaxies in this sample is around solar --$\rm 8.46<12+\log(O/H)<9.22$--, obtained by cross-matching the sample with \cite{2004ApJ...613..898T}. \cite{2017ApJS..233...22S} applied a constant Galactic $\rm \alpha_{CO}=3.2\,M_{\odot}\,(K\,km\,s^{-1}\,pc^{2})^{-1}$ conversion factor that does not include any correction for the presence of helium. The angular size of the galaxies are small enough to fit almost completely inside the IRAM 30 m telescope beam width. Only a small aperture correction of a mean $f_{\rm ap}\sim1.17$ is required. For the aperture correction, they followed the procedure defined in \cite{2011A&A...534A.102L}, as we did for the CO-VGS, with a difference in the exponential scale. They considered an exponential $\rm H_2$ distribution with a half-light radius corresponding to the radius enclosing  50\%  of  the  star  formation  as  measured  in  the SDSS/GALEX photometry, whereas we describe the exponential distribution of the $\rm H_2$ with the exponential scale factor $r_{\rm e}=0.2\times r_{25}$. We do not expect this relatively small  difference to have any impact on our results because the aperture corrections, especially for the CO-VGS sample, are small. Tables \ref{tab:rms_comp} and \ref{tab:ico_comp} compare the $rms$ and the $I_{\rm CO}$ between xCOLD GASS and the CO-VGS; the detection levels are similar.
   
        \begin{table}
            \centering
            \caption{Comparison of the $rms$ between CO-VGS and xCOLD GASS.\label{tab:rms_comp}}
            \begin{tabular}{|c|ccc|ccc|}
                \multicolumn{7}{c}{$rms$ [mK]}\\
                \hline
                \parbox[t]{10mm}{\multirow{2}{*}{\rotatebox[origin=c]{0}{Sample}}}&\multicolumn{3}{c|}{Detections}&\multicolumn{3}{c|}{Non-Detections}\\
                &min&max&mean&min&max&mean\\
                \hline
                \hline
                CO-VGS &1.2 &2.6&1.7&1.0&1.5&1.3\\
                xCOLD GASS & 0.8&4.2&1.8&0.7&3.2&1.3\\
                \hline
            \end{tabular}
        \end{table}
   
        \begin{table}
            \centering
            \caption{Comparison of the $I_{\rm CO}$ between CO-VGS and xCOLD GASS.\label{tab:ico_comp}}
            \begin{tabular}{|c|ccc|}
                \multicolumn{4}{c}{$I_{\rm CO} [{\rm K\,km\,s^{-1}}]$}\\
                \hline
                \parbox[t]{10mm}{\multirow{2}{*}{\rotatebox[origin=c]{0}{Sample}}}&\multicolumn{3}{c|}{Detections}\\
                &min&max&mean\\
                \hline
                \hline
                CO-VGS & 0.5&3.2&1.4\\
                xCOLD GASS &0.1&17.8&1.7\\
                \hline
            \end{tabular}
        \end{table}

        From the xGASS (1179 galaxies), we removed the 333 void galaxies contained in \cite{2012MNRAS.421..926P} and the 69 galaxies classified in \cite{2017A&A...602A.100T} as cluster galaxies. Then, we selected 311 galaxies for the CCS (with H{\tiny I} data and lying within the $M_\star$ and $g-r$ colour ranges of the VGS). For the CO-CS, we selected 95 galaxies (with $\rm H_2$ data and lying within the $M_\star$, $g-r$ colour and SFR ranges of the CO-VGS) from xCOLD GASS and xGASS after removing void and cluster galaxies. Hereafter, we refer to the control galaxies selected from xGASS and xCOLD GASS as GASS galaxies.
   
    \subsubsection{EDGE-CALIFA survey}
    
        EDGE-CALIFA \citep{2017ApJ...846..159B} is a CARMA $\rm H_2$ survey of 126 galaxies selected from the CALIFA \citep{2012A&A...538A...8S} survey that have high WISE $\rm 12\, \mu m$ flux and are centred around 12 hours of right ascension. CALIFA is a diameter-selected survey ($ 45\,{\rm arcsec}<D_{25}<80\,{\rm arsec}$) of 600 galaxies in the redshift range of $\rm 0.005<z<0.030$ and the $M_\star$ range of ${\rm 10^{9.4}\,M_{\odot}}<M_\star<{\rm 10^{11.4}\,M_{\odot}}$. The metallicity is higher than solar ($\rm 8.71<12+\log(O/H)< 9.25$). In EDGE-CALIFA, $M_{\rm H_2}$ was derived considering a constant Galactic $\rm CO$-to-$\rm H_2$ conversion factor $\rm \alpha_{CO}=4.6\,M_{\odot}\,(K\,km\,s^{-1}\,pc^{2})^{-1}$, including the mass correction for the presence of helium. In this work, we rescaled the molecular gas mass of the EDGE-CALIFA galaxies considering $\rm \alpha_{CO}=3.2\,M_{\odot}\,(K\,km\,s^{-1}\,pc^{2})^{-1}$, not including the helium correction, for a consistent comparison with the other surveys.
   
        CARMA is an interferometer and is therefore not sensitive to emission above a certain spatial scale, which means that an extended flux component can be missed. \cite{2017ApJ...846..159B} compared galaxies observed by both the CARMA interferometer and the single-dish IRAM 30 m telescope and concluded that missing flux is not an important problem in the CARMA EDGE-CALIFA observations.
      
        We used H{\tiny I} data from López-Sánchez et al. (in prep.), who  searched the literature for H{\tiny I} data for CALIFA galaxies and found  valid H{\tiny I} data  for 511 objects, the large majority coming from single-dish observations.  Most of the data come from three large surveys: \cite{2005ApJS..160..149S} (305 objects, 60\%), \cite{1989A&A...210....1H} (95 objects, 19\%), and \cite{2004yCat..34300373T} (39 objects, 8\%). We call this sample HI-CALIFA. The remaining galaxies come from 27 references that each provide HI data for between 1 and 14 objects.
  
        For the CCS selection (which does not require CO data), we started from the entire HI-CALIFA sample (511 galaxies) and removed 239 objects with no data in  MPA-JHU, 91 void galaxies contained in \cite{2012MNRAS.421..926P}, and 8 galaxies classified in \cite{2017A&A...602A.100T} as cluster galaxies. Finally, we selected the objects that lie within the VGS colour and stellar mass ranges. We obtained 51 galaxies for the CCS. For the CO-CS, we started from these 51 objects and selected the galaxies with CO data in EDGE-CALIFA  that lie within the CO-VGS colour, stellar mass, and SFR ranges. We  obtained 7 objects for the CO-CS. Hereafter, we refer to this sample as EDGE galaxies. 
    
\section{Results}
    \label{sec:results}
   
    In this section, we compare the gas mass, star formation rate, and stellar mass of void galaxies to those of the control sample. We did this for different sub-samples.
    The CCS was compared with the entire VGS for properties not involving CO, such as $M_{\rm H{\scriptscriptstyle  I}}$ and specific star formation rate (${\rm sSFR= SFR}/M_{\star}$). The CO-VGS was compared with the CO-CS for properties related to CO emission lines such as $M_{\rm H_2}$, molecular-to-atomic gas mass ratio ($M_{\rm H_2}/M_{\rm H{\scriptscriptstyle  I}}$), and SFE, which are not available for the entire VGS. Furthermore, we present all comparisons for the entire samples and also for sub-samples containing only star-forming (SF) galaxies which are close to  the SF main sequence (SFMS). The reason for this limitation is that the VGS galaxies are, partly by selection, mainly SF galaxies and only a few of them fall below the  SFMS. The control sample, on the other hand, contains many galaxies with a very low sSFR (see Fig. \ref{fig:ssfr-mstar_trunc}-left). In order to compare the same type of objects, we excluded quiescent galaxies that are situated well below the SFMS. We adopted the prescription of the SFMS derived  in \cite{2016MNRAS.462.1749S} (their eq. 5) for the COLD GASS sample and derived from it a main sequence in the sSFR ($\rm sSFR_{MS}$) by division with $M_\star$. For our SF sub-sample we then selected the objects that are above the limit $\rm \log_{10}(sSFR/sSFR_{MS})\geq-0.8$, which is represented as the dashed line in Fig. \ref{fig:ssfr-mstar_trunc}. In this way, we removed very low star-forming galaxies from the VGS and from the control sample. We call this selection the SF sub-sample. Additionally, we selected spiral galaxies using the morphological parameter, $\rm t>0$, from HyperLEDA and performed the entire analysis for the spiral sub-sample. We obtained consistent results for the SF and the spiral sub-sample.
   
    We defined three mass bins to compare the CO-VGS and CO-CS as a function of stellar mass: \makebox{$10^{9.0}{\rm M_{\odot}}\leq M_\star<10^{9.5}{\rm M_{\odot}}$}, \makebox{$10^{9.5}{\rm M_{\odot}}\leq M_\star<10^{10.0}{\rm M_{\odot}}$}, \makebox{$10^{10.0}{\rm M_{\odot}}\leq M_\star\leq 10^{10.5}{\rm M_{\odot}}$}, and, in addition, the entire mass range, \makebox{$10^{9.0}{\rm M_{\odot}}\leq M_\star\leq10^{10.5}{\rm M_{\odot}}$}. For the VGS and the CCS, we defined two additional mass bins: \makebox{$10^{8.0}{\rm M_{\odot}}\leq M_\star<10^{8.5}{\rm M_{\odot}}$}, and \makebox{$10^{8.5}{\rm M_{\odot}}\leq M_\star< 10^{9.0}{\rm M_{\odot}}$}, and the entire mass range, \makebox{$10^{8.0}{\rm M_{\odot}}\leq M_\star\leq10^{10.5}{\rm M_{\odot}}$}. We then calculated the mean and median values in every stellar mass bin for both the VGS and the control sample. There are many non-detections for the molecular and atomic gas mass. In order to keep the high statistics, we used the Kaplan-Meier estimator \citep{doi:10.1080/01621459.1958.10501452}, which calculates the mean value taking upper limits into account. 
    As an additional test, we applied the Kolmogorov-Smirnov (KS) test in every stellar bin\footnote{The KS test evaluates whether two samples come from the same mother sample. A p-value below 0.05 indicates with a reliability higher than 95\% that both samples come from different mother samples, whereas for higher p-values, no firm conclusions can be drawn.}
    considering upper limits as detections. The KS test indicates only marginal differences between the samples when the p-value $\lesssim 0.05$, but it denotes high contrast for much lower p-values. We show the corresponding values and the difference between the VGS and the control sample in Tables \ref{tab:mmol-mstar_trunc}-\ref{tab:mmol_matomic-mstar_trunc}.
    
    \subsection{Specific star formation rate}

\begin{figure*}
            \centering
            \includegraphics[width=\hsize]{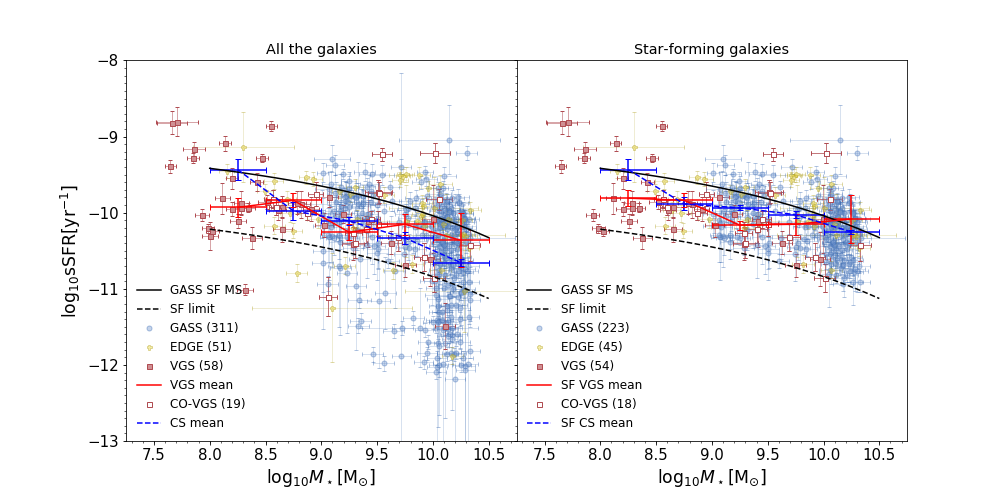}
            \caption{Specific star formation rate as a function of stellar mass for the VGS and CCS with all the galaxies (left), and only star-forming galaxies (right). The mean sSFR per $M_\star$ bin is shown with a red symbol (connected by a solid red line to guide the eye) for the VGS, and with a blue symbol (and dashed blue line) for the CCS. The error bar in $M_\star$ represents the width of the stellar mass bin. The GASS sSFR main sequence is represented as solid black line. This is a fit to the main-sequence galaxies carried out by \cite{2017ApJS..233...22S}. The dashed black line is the limit chosen by us to select star-forming galaxies (see Sect.~\ref{sec:results}).
            }
            \label{fig:ssfr-mstar_trunc}
        \end{figure*}

        The specific star formation rate (${\rm sSFR=SFR}/M_{\star}$, Figure \ref{fig:ssfr-mstar_trunc} and  Table \ref{tab:ssfr-mstar_trunc}) shows a decreasing trend with $M_\star$ for the comparison sample and the VGS, even though it is more pronounced for the comparison sample. %For the entire sample, the mean values of the VGS are above the CCS for stellar masses above $10^{9.5}{\rm M_{\odot}}$. 
        The mean values of both samples lie below the main sequence that was fitted by \cite{2016MNRAS.462.1749S}, which is expected because this fit was made by taking only the star-forming ridge of galaxies into account. It therefore excluded passive galaxies with a low sSFR. Therefore, the agreement between the mean values of the CCS and the main sequence fit is much better for our SF sub-sample (right panel). 
        Interestingly, in this case, there is no significant difference between the void and the comparison sample, except for stellar masses between $10^{9.0}$ and $10^{9.5}{\rm M_{\odot}}$ , where the mean sSFR of the VGS is lower than the CCS ($|\sigma$| > 3 and KS p-value $<0.05$), and the lowest mass bin, where the CCS only contains 3 objects, however.

        \begin{table*}
            \small
            \caption{\label{tab:ssfr-mstar_trunc}Specific star formation rate.}
            \centering
            \begin{tabular}{|ccc|ccccc|ccccc|ccccc|}     % 7 columns 
                \multicolumn{18}{c}{$\rm \log_{10}sSFR[yr^{-1}]$}\\
                \hline
                &\multicolumn{2}{c|}{$\log_{10}M_\star[\rm{M_{\odot}}]$}  &  \multicolumn{5}{c|}{VGS} &  \multicolumn{5}{c|}{CCS} &\multicolumn{5}{c|}{VGS - CCS}\\
                &\multicolumn{2}{c|}{range}  &  $\rm n/n_{up}$ &  \multicolumn{3}{c}{mean} &  median &  $\rm n/n_{up}$ &  \multicolumn{3}{c}{mean} &  median &\multicolumn{3}{c}{$\rm{ \Delta mean}$}&$\sigma$&KS\\
                &\multicolumn{2}{c|}{(1)}  &  (2) &  \multicolumn{3}{c}{(3)} &  (4) &  (5) &  \multicolumn{3}{c}{(6)} &  (7) &\multicolumn{3}{c}{(8)}&(9)&(10)\\
                \hline\hline 
                \parbox[t]{1mm}{\multirow{6}{*}{\rotatebox[origin=c]{90}{ALL}}}
                &   8.0 &   8.5 &      14/1 &    -9.93 &         $\pm$ &         0.12 &      -9.94 &      3/0 &   -9.43 &        $\pm$ &        0.14 &     -9.45 &      -0.50 &           $\pm$ &           0.18 &            - &       - \\
                &   8.5 &   9.0 &      12/0 &    -9.84 &         $\pm$ &         0.09 &      -9.89 &     10/0 &   -9.98 &        $\pm$ &        0.12 &     -9.96 &       0.14 &           $\pm$ &           0.15 &             0.94 &       0.23 \\
                &   9.0 &   9.5 &      10/0 &   -10.26 &         $\pm$ &         0.10 &     -10.16 &     85/0 &  -10.11 &        $\pm$ &        0.05 &    -10.02 &      -0.15 &           $\pm$ &           0.12 &            -1.27 &       0.06 \\
                &   9.5 &  10.0 &      10/1 &   -10.15 &         $\pm$ &         0.14 &     -10.14 &     69/0 &  -10.33 &        $\pm$ &        0.08 &    -10.17 &       0.18 &           $\pm$ &           0.16 &             1.11 &       0.71 \\
                &  10.0 &  10.5 &       5/0 &   -10.36 &         $\pm$ &         0.35 &     -10.43 &    195/0 &  -10.66 &        $\pm$ &        0.05 &    -10.49 &       0.29 &           $\pm$ &           0.36 &             0.82 &       0.50 \\
                \rowcolor{lightgray}
                &   8.0 &  10.5 &      51/2 &   -10.06 &         $\pm$ &         0.07 &      -10.0 &    362/0 &  -10.44 &        $\pm$ &        0.04 &    -10.28 &       0.38 &           $\pm$ &           0.08 &             4.91 &        $2\times 10^{-3}$ \\
                \hline         
                \parbox[t]{1mm}{\multirow{6}{*}{\rotatebox[origin=c]{90}{SF}}}
                &   8.0 &   8.5 &      12/0 &    -9.81 &         $\pm$ &         0.10 &      -9.87 &      3/0 &   -9.43 &        $\pm$ &        0.14 &     -9.45 &      -0.38 &           $\pm$ &           0.17 &            - &       - \\
                &   8.5 &   9.0 &      12/0 &    -9.84 &         $\pm$ &         0.09 &      -9.89 &      9/0 &   -9.88 &        $\pm$ &        0.09 &     -9.96 &       0.05 &           $\pm$ &           0.12 &             0.38 &       0.42 \\
                &   9.0 &   9.5 &       9/0 &   -10.16 &         $\pm$ &         0.06 &     -10.16 &     71/0 &   -9.94 &        $\pm$ &        0.03 &     -9.92 &      -0.23 &           $\pm$ &           0.07 &            -3.33 &       0.02 \\
                &   9.5 &  10.0 &      10/0 &   -10.15 &         $\pm$ &         0.14 &     -10.14 &     54/0 &  -10.03 &        $\pm$ &        0.05 &    -10.04 &      -0.12 &           $\pm$ &           0.15 &            -0.84 &       0.54 \\
                &  10.0 &  10.5 &       4/0 &   -10.08 &         $\pm$ &         0.31 &      -9.82 &    131/0 &  -10.25 &        $\pm$ &        0.03 &    -10.22 &       0.17 &           $\pm$ &           0.31 &             0.55 &       0.38 \\
                \rowcolor{lightgray}
                &   8.0 &  10.5 &      47/0 &    -9.98 &         $\pm$ &         0.06 &      -9.95 &    268/0 &   -10.10 &        $\pm$ &        0.02 &     -10.10 &       0.12 &           $\pm$ &           0.06 &             1.93 &       0.07 \\
                \hline
            \end{tabular}
            \tablefoot{(1) Stellar mass range of the bin. (2) n: Number of VGS galaxies in the bin. $\rm n_{up}$: Number of upper limits of VGS galaxies in the bin. (3) Mean logarithm of the specific star formation rate and its error of the VGS galaxies in the bin. (4) Median logarithm of the specific star formation rate of the VGS galaxies in the bin. (5) - (7) The same for the CCS sample. (8) Difference of the mean logarithmic of the specific star formation rate between VGS and CCS ($\rm{ \Delta mean}$) and its error ($\rm err( \Delta mean)$). (9) $\rm \sigma = { \Delta mean}/err({\Delta mean})$,  only reported when there are at least four objects in each sample. (10) p-value of the Kolmogorov-Smirnov test.} 
        \end{table*}
 
    \subsection{Molecular gas mass} \label{sec:molecular}
   
\begin{figure*}
            \centering
            \includegraphics[width=\hsize]{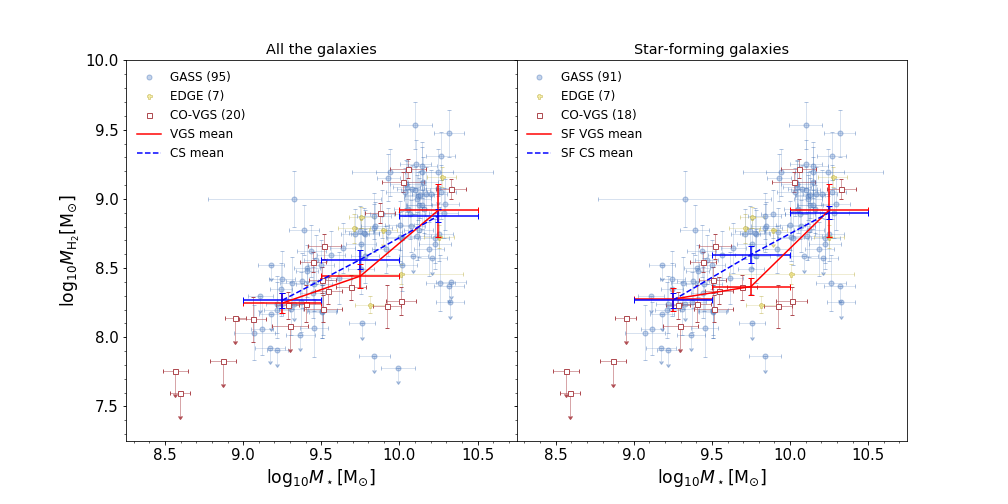}
            \caption{Molecular gas mass as a function of stellar mass for the CO-VGS and CO-CS with all the galaxies (left) and only star-forming galaxies (right). The mean $M_{\rm H_2}$ per $M_\star$ bin is shown with a red symbol (connected by a solid red line to guide the eye) for the CO-VGS, and with a blue symbol (and dashed blue line) for the CO-CS. The error bar in $M_\star$ represents the width of the stellar mass bin.
            }
            \label{fig:mmol-mstar_trunc}
        \end{figure*}

        The molecular gas mass shows an increasing trend with $M_{\star}$ for the void and the comparison samples (Fig. \ref{fig:mmol-mstar_trunc} and Table~\ref{tab:mmol-mstar_trunc}). In general, the mean values of the CO-VGS and CO-CS samples agree within the errors, except for the intermediate $M_\star$ bin, where the mean $M_{\rm H_2}$ for void galaxies is slightly lower  than for galaxies in filaments and walls. The difference is marginal ($|\sigma| \sim 1$) when the entire samples are considered and is slighly larger for the SF sub-sample ($|\sigma|\sim 2.6$ and KS p-value $<0.05$). 
 
        \begin{table*}
            \small
            \caption{\label{tab:mmol-mstar_trunc} Comparison of molecular gas mass between CO-VGS and CO-CS.}
            \centering
            \begin{tabular}{|ccc|ccccc|ccccc|ccccc|}     % 7 columns 

                \multicolumn{18}{c}{$\log_{10}M_{\rm H_2}[\rm{M_{\odot}}]$}\\
                \hline
                &\multicolumn{2}{c|}{$\log_{10}M_\star[{\rm M_{\odot}}]$}  &  \multicolumn{5}{c|}{CO-VGS} &  \multicolumn{5}{c|}{CO-CS} &\multicolumn{5}{c|}{CO-VGS - CO-CS}\\
                &\multicolumn{2}{c|}{range}  &  $\rm n/n_{up}$ &  \multicolumn{3}{c}{mean} &  median &  $\rm n/n_{up}$ &  \multicolumn{3}{c}{mean} &  median &\multicolumn{3}{c}{$\rm{ \Delta mean}$}& $\sigma$&KS\\
                &\multicolumn{2}{c|}{(1)}  &  (2) &  \multicolumn{3}{c}{(3)} &  (4) &  (5) &  \multicolumn{3}{c}{(6)} &  (7) &\multicolumn{3}{c}{(8)}&(9)&(10)\\
                \hline \hline 
                \parbox[t]{1mm}{\multirow{4}{*}{\rotatebox[origin=c]{90}{ALL}}}
                &   9.0 &   9.5 &       5/1 &     8.25 &         $\pm$ &         0.07 &       8.23 &     26/8 &    8.27 &        $\pm$ &        0.06 &      8.24 &      -0.02 &           $\pm$ &           0.09 &            -0.23 &       0.48 \\
                &   9.5 &  10.0 &       7/0 &     8.44 &         $\pm$ &         0.09 &       8.36 &     29/4 &    8.56 &        $\pm$ &        0.07 &      8.64 &      -0.12 &           $\pm$ &           0.11 &            -1.07 &       0.11 \\
                &  10.0 &  10.5 &       4/0 &     8.92 &         $\pm$ &         0.19 &       9.12 &     47/6 &    8.88 &        $\pm$ &        0.05 &      8.93 &       0.04 &           $\pm$ &           0.20 &             0.21 &       0.23 \\
                \rowcolor{lightgray}
                &   9.0 &  10.5 &      16/1 &      8.50 &         $\pm$ &         0.09 &       8.36 &   102/18 &    8.61 &        $\pm$ &        0.04 &      8.72 &      -0.11 &           $\pm$ &            0.10 &            -1.14 &       0.06 \\
                \hline        
                \parbox[t]{1mm}{\multirow{4}{*}{\rotatebox[origin=c]{90}{SF}}}
                &   9.0 &   9.5 &       4/1 &     8.28 &         $\pm$ &         0.08 &       8.24 &     26/8 &    8.27 &        $\pm$ &        0.06 &      8.24 &       0.01 &           $\pm$ &           0.10 &             0.08 &       0.76 \\
                &   9.5 &  10.0 &       6/0 &     8.37 &         $\pm$ &         0.06 &       8.34 &     28/3 &    8.60 &        $\pm$ &        0.06 &      8.68 &      -0.23 &           $\pm$ &           0.09 &            -2.64 &       0.02 \\
                &  10.0 &  10.5 &       4/0 &     8.92 &         $\pm$ &         0.19 &       9.12 &     44/4 &    8.90 &        $\pm$ &        0.05 &      8.95 &       0.02 &           $\pm$ &           0.20 &             0.09 &       0.22 \\
                \rowcolor{lightgray}
                &   9.0 &  10.5 &      14/1 &      8.50 &         $\pm$ &          0.10 &       8.34 &    98/15 &    8.64 &        $\pm$ &        0.04 &      8.73 &      -0.14 &           $\pm$ &           0.11 &            -1.33 &       0.05 \\
                \hline
                \end{tabular}
            \tablefoot{(1) Stellar mass range of the bin. (2) Number of CO-VGS galaxies in the bin.  $\rm n_{up}$: Number of upper limits of CO-VGS galaxies in the bin. (3) Mean logarithm of the molecular gas mass and its error of the CO-VGS galaxies in the bin. (4) Median logarithm of the molecular gas mass of the CO-VGS galaxies in the bin. (5) - (7) The same for the CO-CS sample. (8) Difference of the mean logarithmic of the molecular gas mass between CO-VGS and CO-CS ($\rm{ \Delta mean}$) and its error ($\rm err( \Delta mean)$). (9) $\rm \sigma = { \Delta mean}/err({\Delta mean})$. (10) p-value of the Kolmogorov-Smirnov test.
            }
        \end{table*}

\begin{figure*}
            \centering
            \includegraphics[width=\hsize]{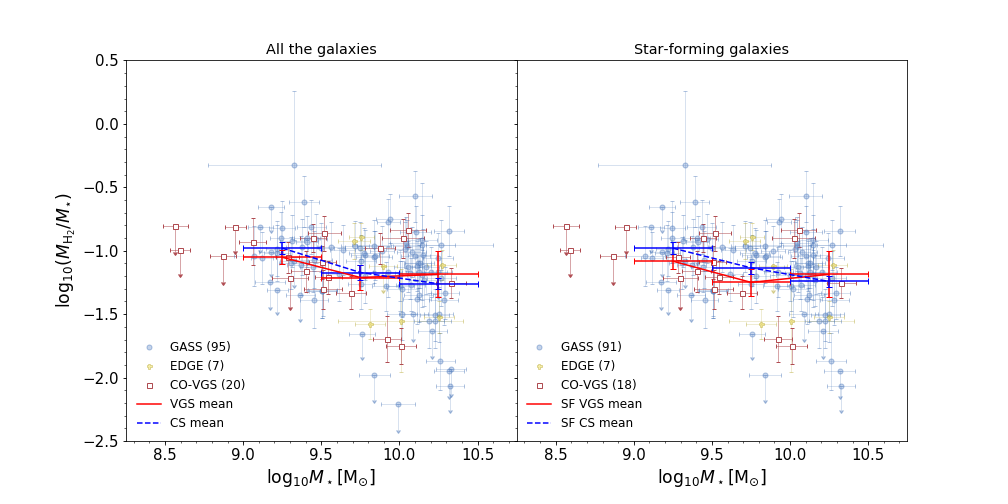}
            \caption{Molecular gas mass fraction as a function of stellar mass for the CO-VGS and CO-CS with all the galaxies (left) and only star-forming galaxies (right). The mean $M_{\rm H_2}/M_\star$ per $M_\star$ bin is shown with a red symbol (connected by a solid red line to guide the eye) for the CO-VGS, and with a blue symbol (and dashed blue line) for the CO-CS. The error bar in $M_\star$ represents the width of the stellar mass bin.
            }
            \label{fig:fracmol-mstar_trunc}
        \end{figure*}

        The molecular gas mass fraction ($M_{\rm H_2}/M_\star$) shows a decreasing trend with $M_{\star}$ for the void and the comparison samples (Fig. \ref{fig:fracmol-mstar_trunc} and Table \ref{tab:fracmol-mstar_trunc}). There is no significant difference between the two samples ($|\sigma| < 1$) for the complete sample or for the SF sub-sample. 

        In summary, we conclude that we find no significant difference for $M_{\rm H_2}$ or $M_{\rm H_2}/M_\star$ between CO-VGS and CO-CS. For both samples, we find decreasing trends of $M_{\rm H_2}/M_\star$ with $M_\star$.

        \begin{table*}
            \small
            \caption{\label{tab:fracmol-mstar_trunc}Molecular gas mass fraction.}
            \centering
            \begin{tabular}{|ccc|ccccc|ccccc|ccccc|}     % 7 columns 
                \multicolumn{18}{c}{$\log_{10}(M_{\rm H_2}/M_\star$)}\\
                \hline
                &\multicolumn{2}{c|}{$\log_{10}M_\star[{\rm M_{\odot}}]$}  &  \multicolumn{5}{c|}{CO-VGS} &  \multicolumn{5}{c|}{CO-CS} &\multicolumn{5}{c|}{CO-VGS - CO-CS}\\
                &\multicolumn{2}{c|}{range}  &  $\rm n/n_{up}$ &  \multicolumn{3}{c}{mean} &  median &  $\rm n/n_{up}$ &  \multicolumn{3}{c}{mean} &  median &\multicolumn{3}{c}{$\rm{ \Delta mean}$}&$\sigma$&KS\\
                &\multicolumn{2}{c|}{(1)}  &  (2) &  \multicolumn{3}{c}{(3)} &  (4) &  (5) &  \multicolumn{3}{c}{(6)} &  (7) &\multicolumn{3}{c}{(8)}&(9)&(10)\\
                \hline\hline 
                \parbox[t]{1mm}{\multirow{4}{*}{\rotatebox[origin=c]{90}{ALL}}}
                &   9.0 &   9.5 &       5/1 &    -1.05 &         $\pm$ &         0.06 &      -1.05 &     26/8 &   -1.04 &        $\pm$ &        0.05 &     -1.04 &      -0.01 &           $\pm$ &           0.08 &            -0.16 &       0.87 \\
                &   9.5 &  10.0 &       7/0 &    -1.21 &         $\pm$ &         0.10 &      -1.21 &     29/4 &   -1.21 &        $\pm$ &        0.07 &     -1.05 &       0.00 &           $\pm$ &           0.12 &             0.01 &       0.75 \\
                &  10.0 &  10.5 &       4/0 &    -1.19 &         $\pm$ &         0.18 &      -0.90 &     47/6 &   -1.29 &        $\pm$ &        0.05 &     -1.19 &       0.10 &           $\pm$ &           0.19 &             0.54 &       0.37 \\
                \rowcolor{lightgray}
                &   9.0 &  10.5 &      16/1 &    -1.17 &         $\pm$ &         0.07 &      -1.09 &   102/18 &   -1.23 &        $\pm$ &        0.04 &     -1.12 &       0.06 &           $\pm$ &           0.08 &             0.67 &       0.97 \\
                \hline         
                \parbox[t]{1mm}{\multirow{4}{*}{\rotatebox[origin=c]{90}{SF}}}
                &   9.0 &   9.5 &       4/1 &    -1.08 &         $\pm$ &         0.06 &      -1.05 &     26/8 &   -1.04 &        $\pm$ &        0.05 &     -1.04 &      -0.04 &           $\pm$ &           0.08 &            -0.54 &       0.63 \\
                &   9.5 &  10.0 &       6/0 &    -1.25 &         $\pm$ &         0.10 &      -1.30 &     28/3 &   -1.16 &        $\pm$ &        0.06 &     -1.04 &      -0.09 &           $\pm$ &           0.12 &            -0.78 &       0.31 \\
                &  10.0 &  10.5 &       4/0 &    -1.19 &         $\pm$ &         0.18 &      -0.90 &     44/4 &   -1.26 &        $\pm$ &        0.05 &     -1.19 &       0.07 &           $\pm$ &           0.19 &             0.38 &       0.39 \\
                \rowcolor{lightgray}
                &   9.0 &  10.5 &      14/1 &     -1.20 &         $\pm$ &         0.08 &      -1.21 &    98/15 &   -1.19 &        $\pm$ &        0.04 &     -1.12 &      -0.01 &           $\pm$ &           0.09 &            -0.17 &       0.82 \\
                \hline
            \end{tabular}
            \tablefoot{(1) Stellar mass range of the bin. (2) Number of CO-VGS galaxies in the bin. $\rm n_{up}$: Number of upper limits of CO-VGS galaxies in the bin. (3) Mean logarithm of the molecular gas mass fraction and its error of the CO-VGS galaxies in the bin. (4) Median logarithm of the molecular gas mass fraction of the CO-VGS galaxies in the bin. (5) - (7) The same for the CO-CS sample. (8) Difference of the mean logarithmic of the molecular gas mass fraction between CO-VGS and CO-CS ($\rm{ \Delta mean}$) and its error ($\rm err( \Delta mean)$). (9) $\rm \sigma = { \Delta mean}/err({\Delta mean})$. (10) p-value of the Kolmogorov-Smirnov test applied inside the bin to compare the CO-VGS and the CO-CS.
            }
        \end{table*}
    
    \subsection{Star formation efficiency}
 
\begin{figure*}
            \centering
            \includegraphics[width=\hsize]{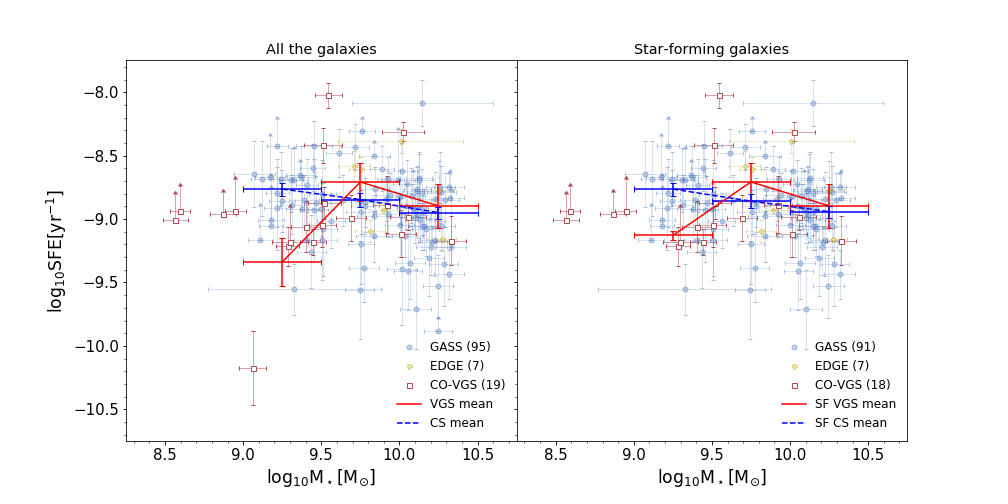}
            \caption{Star formation efficiency as a function of stellar mass for the CO-VGS and CO-CS with all the galaxies (left) and only star-forming galaxies (right). The mean SFE per $M_\star$ bin is shown with a red symbol (connected by a solid red line to guide the eye) for the CO-VGS, and with a blue symbol (and dashed blue line) for the CO-CS. The error bar in $M_\star$ represents the width of the stellar mass bin.
            }
            \label{fig:sfe-mstar_trunc}
        \end{figure*}

        The star formation efficiency (${\rm SFE=SFR}/M_{\rm H_2}$, Fig. \ref{fig:sfe-mstar_trunc} and Table~\ref{tab:sfe-mstar_trunc}) shows a  decreasing trend with $M_\star$ for the comparison sample. This trend is followed by the void galaxies for the intermediate and upper $M_\star$ bin, where the mean values agree within 1 $\sigma$. However, in the lowest-mass bin, void galaxies have a significantly ($|\sigma| \gtrsim 3$ and KS p-value $\sim 0.002$) lower mean SFE for the entire sample and for the SF sub-sample. The number of galaxies in this bin is relatively low (four to six galaxies), and it needs to be confirmed with a larger sample size. 

        \begin{table*}
            \small
            \caption{\label{tab:sfe-mstar_trunc}Star formation efficiency.}
            \centering
            \begin{tabular}{|ccc|ccccc|ccccc|ccccc|}     % 7 columns 
                \multicolumn{18}{c}{$\rm \log_{10}SFE[yr^{-1}]$}\\
                \hline
                &\multicolumn{2}{c|}{$\log_{10}M_\star[\rm{M_{\odot}}]$}  &  \multicolumn{5}{c|}{CO-VGS} &  \multicolumn{5}{c|}{CO-CS} &\multicolumn{5}{c|}{CO-VGS - CO-CS}\\
                &\multicolumn{2}{c|}{range}  &  $\rm n/n_{up}$ &  \multicolumn{3}{c}{mean} &  median &  $\rm n/n_{up}$ &  \multicolumn{3}{c}{mean} &  median &\multicolumn{3}{c}{$\rm{ \Delta mean}$}&$\rm{\sigma}$&KS\\
                &\multicolumn{2}{c|}{(1)}  &  (2) &  \multicolumn{3}{c}{(3)} &  (4) &  (5) &  \multicolumn{3}{c}{(6)} &  (7) &\multicolumn{3}{c}{(8)}&(9)&(10)\\
                \hline\hline 
                \parbox[t]{1mm}{\multirow{4}{*}{\rotatebox[origin=c]{90}{ALL}}}
                &   9.0 &   9.5 &       5/0 &    -9.34 &         $\pm$ &         0.19 &      -9.18 &     26/0 &   -8.76 &        $\pm$ &        0.05 &     -8.69 &      -0.57 &           $\pm$ &           0.20 &            -2.93 &        $2\times 10^{-3}$ \\
                &   9.5 &  10.0 &       6/1 &    -8.71 &         $\pm$ &         0.15 &      -8.87 &     29/0 &   -8.85 &        $\pm$ &        0.06 &     -8.85 &       0.14 &           $\pm$ &           0.16 &             0.87 &       0.27 \\
                &  10.0 &  10.5 &       4/0 &    -8.90 &         $\pm$ &         0.17 &      -9.12 &     47/0 &   -8.95 &        $\pm$ &        0.05 &     -8.90 &       0.05 &           $\pm$ &           0.18 &             0.30 &       0.88 \\
                \rowcolor{lightgray}
                &   9.0 &  10.5 &      15/1 &    -8.95 &         $\pm$ &         0.13 &      -8.99 &    102/0 &   -8.86 &        $\pm$ &        0.03 &     -8.85 &      -0.09 &           $\pm$ &           0.13 &            -0.68 &       0.22 \\
                \hline         
                \parbox[t]{1mm}{\multirow{4}{*}{\rotatebox[origin=c]{90}{SF}}}
                &   9.0 &   9.5 &       4/0 &    -9.13 &         $\pm$ &         0.04 &      -9.18 &     26/0 &   -8.76 &        $\pm$ &        0.05 &     -8.69 &      -0.37 &           $\pm$ &           0.06 &            -5.85 &       0.01 \\
                &   9.5 &  10.0 &       6/0 &    -8.71 &         $\pm$ &         0.15 &      -8.87 &     28/0 &   -8.86 &        $\pm$ &        0.06 &     -8.87 &       0.15 &           $\pm$ &           0.16 &             0.95 &       0.69 \\
                &  10.0 &  10.5 &       4/0 &    -8.90 &         $\pm$ &         0.17 &      -9.12 &     44/0 &   -8.94 &        $\pm$ &        0.05 &     -8.90 &       0.05 &           $\pm$ &           0.18 &             0.25 &       0.77 \\
                \rowcolor{lightgray}
                &   9.0 &  10.5 &      14/0 &    -8.86 &         $\pm$ &          0.10 &      -8.99 &     98/0 &   -8.86 &        $\pm$ &        0.03 &     -8.85 &       -0.00 &           $\pm$ &           0.11 &            -0.02 &       0.16 \\
                \hline
            \end{tabular}
            \tablefoot{(1) Stellar mass range of the bin. (2) Number of CO-VGS galaxies in the bin. $\rm n_{up}$: Number of upper limits of CO-VGS galaxies in the bin. (3) Mean logarithm of the star formation efficiency and its error of the CO-VGS galaxies in the bin. (4) Median logarithm of the star formation efficiency of the CO-VGS galaxies in the bin. (5) - (7) The same for the CO-CS sample. (8) Difference of the mean logarithmic of the star formation efficiency between CO-VGS and CO-CS ($\rm{ \Delta mean}$) and its error ($\rm err( \Delta mean)$). (9) $\rm \sigma = { \Delta mean}/err({\Delta mean})$. (10) p-value of the Kolmogorov-Smirnov test.
            }
        \end{table*}

    \subsection{Atomic gas mass}\label{sec:atomic}
    
\begin{figure*}
            \centering
            \includegraphics[width=\hsize]{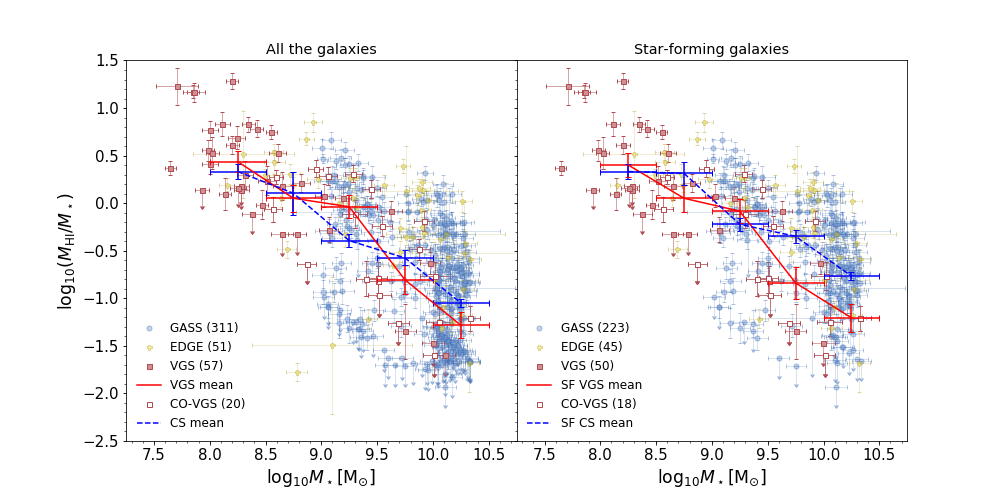}
            \caption{Atomic gas mass fraction as a function of stellar mass for the VGS and CCS with all the galaxies (left) and only star-forming galaxies (right). The mean $M_{\rm H{\scriptscriptstyle  I}}/M_{\star}$ per $M_\star$ bin is shown with a red symbol (connected by a  solid red line to guide the eye) for the VGS, and with a blue symbol (and dashed blue line) for the CCS. The error bar in $M_\star$ represents the width of the stellar mass bin.
            }
            \label{fig:fracatom-mstar_trunc}
        \end{figure*}

        The atomic gas mass fraction ($M_{\rm H{\scriptscriptstyle  I}}/M_{\star}$, Fig. \ref{fig:fracatom-mstar_trunc} and Table \ref{tab:fracatom-mstar_trunc}) shows a strongly decreasing trend with $M_{\star}$ for both the VGS and CCS. We can  directly compare the VGS to the comparison sample for $M_{\star} > 10^{8.0}\, {\rm M_\odot}$, and the trend for the VGS galaxies seems to  follow the trend of the CCS very well.
   
        In general, the mean values of void and control galaxies agree reasonably well, except for galaxies with $M_{\star} > 10^{9.0}\, {\rm M_\odot}$ , which show indications for a steeper slope for the VGS. In the SF sub-sample, the mean $M_{\rm H{\scriptscriptstyle  I}}/M_{\star}$ of the VGS is lower than the CCS for $10^{9.5}{\rm M_{\odot}}<M_{\star}<10^{10.5}{\rm M_{\odot}}$, but the difference is statistically marginal  ($|\sigma| \lesssim 3$ and KS p-value $\gtrsim 0.05$). Furthermore,    for the highest stellar mass bin  ($10^{10.0}{\rm M_{\odot}}<M_{\star}<10^{10.5}{\rm M_{\odot}}$), it is based on a very low number of galaxies (four). 
   
        \begin{table*}
            \small
            \caption{\label{tab:fracatom-mstar_trunc}Atomic gas mass fraction.}
            \centering
            \begin{tabular}{|ccc|ccccc|ccccc|ccccc|}     % 7 columns 
                \multicolumn{18}{c}{$\log_{10}(M_{\rm H{\scriptscriptstyle  I}}/M_\star)$}\\
                \hline
                &\multicolumn{2}{c|}{$\log_{10}M_\star[\rm{M_{\odot}}]$}  &  \multicolumn{5}{c|}{VGS} &  \multicolumn{5}{c|}{CCS} &\multicolumn{5}{c|}{VGS - CCS}\\
                &\multicolumn{2}{c|}{range}  & $\rm n/n_{up}$ &  \multicolumn{3}{c}{mean} &  median &  $\rm n/n_{up}$ &  \multicolumn{3}{c}{mean} &  median &\multicolumn{3}{c}{$\rm{ \Delta mean}$}&$\sigma$&KS\\
                &\multicolumn{2}{c|}{(1)}  &  (2) &  \multicolumn{3}{c}{(3)} &  (4) &  (5) &  \multicolumn{3}{c}{(6)} &  (7) &\multicolumn{3}{c}{(8)}&(9)&(10)\\
                \hline\hline 
                \parbox[t]{1mm}{\multirow{4}{*}{\rotatebox[origin=c]{90}{ALL}}}
                 &   8.0 &   8.5 &      15/4 &     0.44 &         $\pm$ &         0.11 &       0.53 &      3/0 &    0.33 &        $\pm$ &        0.08 &      0.29 &       0.11 &           $\pm$ &           0.13 &             - &       - \\
                &   8.5 &   9.0 &      11/4 &     0.05 &         $\pm$ &         0.14 &       0.21 &     10/0 &    0.10 &        $\pm$ &        0.23 &      0.31 &      -0.05 &           $\pm$ &           0.27 &            -0.20 &       0.98 \\
                &   9.0 &   9.5 &       8/2 &    -0.04 &         $\pm$ &         0.12 &       0.06 &    85/13 &   -0.39 &        $\pm$ &        0.07 &     -0.22 &       0.36 &           $\pm$ &           0.14 &             2.54 &       0.09 \\
                &   9.5 &  10.0 &      11/3 &    -0.81 &         $\pm$ &         0.15 &      -0.79 &    69/12 &   -0.58 &        $\pm$ &        0.08 &     -0.40 &      -0.23 &           $\pm$ &           0.17 &            -1.33 &       0.33 \\
                &  10.0 &  10.5 &       5/2 &    -1.28 &         $\pm$ &         0.14 &      -1.25 &   195/56 &   -1.05 &        $\pm$ &        0.05 &     -0.95 &      -0.23 &           $\pm$ &           0.14 &            -1.62 &       0.24 \\
                \rowcolor{lightgray}
                &   8.0 &  10.5 &     50/15 &    -0.27 &         $\pm$ &         0.12 &      -0.08 &   362/81 &   -0.77 &        $\pm$ &        0.04 &     -0.71 &       0.51 &           $\pm$ &           0.12 &              4.10 &        $6\times 10^{-7}$ \\
                \hline         
                \parbox[t]{1mm}{\multirow{4}{*}{\rotatebox[origin=c]{90}{SF}}}
                &   8.0 &   8.5 &      12/4 &     0.40 &         $\pm$ &         0.13 &       0.53 &      3/0 &    0.33 &        $\pm$ &        0.08 &      0.29 &       0.07 &           $\pm$ &           0.15 &             - &       - \\
                &   8.5 &   9.0 &      11/4 &     0.05 &         $\pm$ &         0.14 &       0.21 &      9/0 &    0.31 &        $\pm$ &        0.12 &      0.31 &      -0.26 &           $\pm$ &           0.19 &            -1.41 &       0.85 \\
                &   9.0 &   9.5 &       7/2 &    -0.08 &         $\pm$ &         0.13 &       0.06 &     71/5 &   -0.22 &        $\pm$ &        0.07 &     -0.06 &       0.14 &           $\pm$ &           0.14 &             0.95 &       0.40 \\
                &   9.5 &  10.0 &      10/3 &    -0.84 &         $\pm$ &         0.17 &      -0.79 &     54/3 &   -0.35 &        $\pm$ &        0.07 &     -0.28 &      -0.49 &           $\pm$ &           0.18 &            -2.75 &       0.03 \\
                &  10.0 &  10.5 &       4/1 &    -1.20 &         $\pm$ &         0.15 &      -1.21 &   131/11 &   -0.77 &        $\pm$ &        0.04 &     -0.72 &      -0.44 &           $\pm$ &           0.15 &            -2.85 &       0.08 \\
                \rowcolor{lightgray}
                &   8.0 &  10.5 &     44/14 &     -0.30 &         $\pm$ &         0.12 &      -0.12 &   268/19 &    -0.50 &        $\pm$ &        0.04 &     -0.44 &        0.20 &           $\pm$ &           0.13 &             1.52 &         $5\times 10^{-4}$ \\
                \hline
            \end{tabular}
            \tablefoot{(1) Stellar mass range of the bin. (2) Number of VGS galaxies in the bin. $\rm n_{up}$: Number of upper limits of VGS galaxies in the bin. (3) Mean logarithm of the atomic gas mass fraction and its error of the VGS galaxies in the bin. (4) Median logarithm of the atomic gas mass fraction of the VGS galaxies in the bin. (5) - (7) The same for the CCS sample. (8) Difference of the mean logarithmic of the atomic gas mass fraction between VGS and CCS ($\rm{ \Delta mean}$) and its error ($\rm err( \Delta mean)$). (9) $\rm \sigma = { \Delta mean}/err({\Delta mean})$, only reported when there are at least 4 objects in each sample. (10) p-value of the Kolmogorov-Smirnov test.
            }
        \end{table*}

    \subsection{Molecular-to-atomic gas mass ratio}

\begin{figure*}
            \centering
            \includegraphics[width=\hsize]{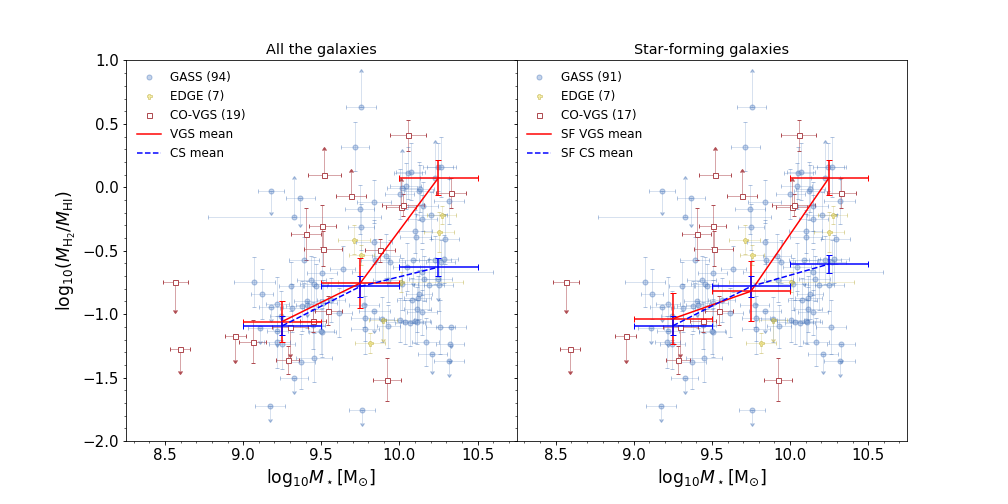}
            \caption{Molecular-to-atomic gas mass ratio as function of stellar mass for the CO-VGS and the CO-CS with all the galaxies (left) and only star-forming galaxies (right). The mean $M_{\rm H_2}/M_{\rm H{\scriptscriptstyle  I}}$ per $M_\star$ bin is calculated taking upper limits into account but not lower ones, and it is shown with a red symbol (connected by a solid red line to guide the eye) for the CO-VGS, and with a blue symbol (and dashed blue line) for the CO-CS. The error bar in $M_\star$ represents the width of the stellar mass bin. 
            }
            \label{fig:mmol_matomic-mstar_trunc}
        \end{figure*}

        The molecular-to-atomic gas mass ratio (Fig. \ref{fig:mmol_matomic-mstar_trunc} and Table \ref{tab:mmol_matomic-mstar_trunc}) shows an increasing trend with $M_\star$ for the void and the comparison samples.
        In the low and intermediate stellar mass bins, the agreement between the mean values of the VGS and comparison sample is very good, whereas in the high stellar mass bin, the mean value for the VGS is considerably higher. However, this difference has to be taken with caution because of the low number of VGS galaxies in this bin.

        Because the Kaplan-Meier estimator can only deal with upper or lower limits but not with both, we included only the upper limits (i.e. upper limits in  $M_{\rm H_2}$ and detections in $M_{\rm H{\scriptscriptstyle  I}}$) in the calculation of the mean. We  also carried out this analysis with only lower limits (i.e. upper limits in  $M_{\rm H{\scriptscriptstyle  I}}$ and detection in $M_{\rm H_2}$) and obtained consistent results.

        \begin{table*}
            \small
            \caption{\label{tab:mmol_matomic-mstar_trunc}Molecular-to-atomic gas mass ratio. Means have been calculated taking upper limits into account, but not lower limits.}
            \centering
            \begin{tabular}{|ccc|ccccc|ccccc|ccccc|}     % 7 columns 
                \multicolumn{18}{c}{$\log_{10}(M_{\rm H_2}/M_{\rm H{\scriptscriptstyle  I}})$ (Upper limits)}\\
                \hline
                &\multicolumn{2}{c|}{$\log_{10}M_\star[\rm{M_{\odot}}]$}  &  \multicolumn{5}{c|}{CO-VGS} &  \multicolumn{5}{c|}{CO-CS} &\multicolumn{5}{c|}{CO-VGS - CO-CS}\\
                &\multicolumn{2}{c|}{range}  &  $\rm n/n_{up}$ &  \multicolumn{3}{c}{mean} &  median &  $\rm n/n_{up}$ &  \multicolumn{3}{c}{mean} &  median &\multicolumn{3}{c}{$\rm{ \Delta mean}$}&$\sigma$&KS\\
                &\multicolumn{2}{c|}{(1)}  &  (2) &  \multicolumn{3}{c}{(3)} &  (4) &  (5) &  \multicolumn{3}{c}{(6)} &  (7) &\multicolumn{3}{c}{(8)}&(9)&(10)\\
                \hline\hline 
                \parbox[t]{1mm}{\multirow{4}{*}{\rotatebox[origin=c]{90}{ALL}}}
                &   9.0 &   9.5 &       5/1 &    -1.06 &         $\pm$ &         0.16 &      -1.22 &     25/8 &   -1.09 &        $\pm$ &        0.07 &     -1.01 &       0.03 &           $\pm$ &           0.18 &             0.19 &       0.26 \\
                &   9.5 &  10.0 &       5/0 &    -0.75 &         $\pm$ &         0.20 &      -0.50 &     27/3 &   -0.78 &        $\pm$ &        0.08 &     -0.89 &       0.03 &           $\pm$ &           0.21 &             0.12 &       0.17 \\
                &  10.0 &  10.5 &       3/0 &     0.08 &         $\pm$ &         0.14 &      -0.05 &     45/6 &   -0.63 &        $\pm$ &        0.07 &     -0.72 &       0.70 &           $\pm$ &           0.15 &             4.57 &       0.02 \\
                \rowcolor{lightgray}
                &   9.0 &  10.5 &      13/1 &    -0.69 &         $\pm$ &         0.16 &       -0.5 &    97/17 &    -0.80 &        $\pm$ &        0.05 &     -0.84 &       0.11 &           $\pm$ &           0.17 &             0.66 &       0.08 \\
                \hline         
                \parbox[t]{1mm}{\multirow{4}{*}{\rotatebox[origin=c]{90}{SF}}}
                &   9.0 &   9.5 &       4/1 &    -1.04 &         $\pm$ &         0.20 &      -1.05 &     25/8 &   -1.09 &        $\pm$ &        0.07 &     -1.01 &       0.06 &           $\pm$ &           0.22 &             0.26 &       0.51 \\
                &   9.5 &  10.0 &       4/0 &    -0.82 &         $\pm$ &         0.23 &      -0.48 &     27/3 &   -0.78 &        $\pm$ &        0.08 &     -0.89 &      -0.04 &           $\pm$ &           0.25 &            -0.15 &       0.27 \\
                &  10.0 &  10.5 &       3/0 &     0.08 &         $\pm$ &         0.14 &      -0.05 &     43/4 &   -0.60 &        $\pm$ &        0.07 &     -0.67 &       0.68 &           $\pm$ &           0.15 &             - &       - \\
                \rowcolor{lightgray}
                &   9.0 &  10.5 &      11/1 &    -0.66 &         $\pm$ &         0.18 &      -0.48 &    95/15 &   -0.79 &        $\pm$ &        0.05 &     -0.84 &       0.13 &           $\pm$ &           0.19 &             0.66 &       0.07 \\
                \hline
            \end{tabular}
            \tablefoot{(1) Stellar mass range of the bin. (2) Number of CO-VGS galaxies in the bin. $\rm n_{up}$: Number of upper limits of CO-VGS galaxies in the bin. (3) Mean logarithm of the molecular-to-atomic gas mass ratio and its error of the CO-VGS galaxies in the bin. (4) Median logarithm of the molecular-to-atomic gas mass ratio of the CO-VGS galaxies in the bin. (5) - (7) The same for the CO-CS sample. (8) Difference of the mean logarithmic of the molecular-to-atomic gas mass ratio between CO-VGS and CO-CS ($\rm{ \Delta mean}$) and its error ($\rm err( \Delta mean)$). (9) $\rm \sigma = { \Delta mean}/err({\Delta mean})$, only reported when there are at least four objects in each sample. (10) p-value of the Kolmogorov-Smirnov test.
            }
        \end{table*}

    \subsection{$\rm CO(2-1)$-to-$\rm CO(1-0)$ line ratio}
        \label{sec:line-ratio}
 
\begin{figure*}
            \centering
            \begin{tabular}{cc}
                \includegraphics[width=9cm]{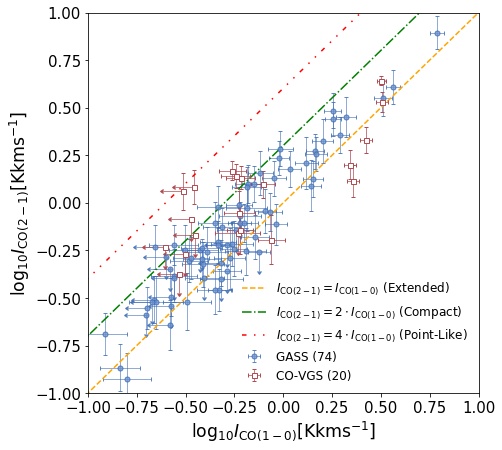}&
                \includegraphics[width=8.4cm]{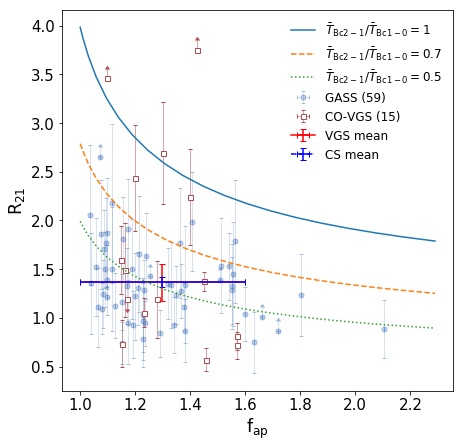}\\
     
            \end{tabular}
            \caption{ Correlation between the CO emission line intensities. (Left) $I_{\rm CO(2-1)}$ and $I_{\rm CO(1-0)}$ emission line comparison for CO-VGS and xCOLD GASS galaxies. (Right) Emission line ratio ($R_{\rm 21}=I_{\rm CO(2-1)}/I_{\rm CO(1-0)}$) as a function of the aperture-correction factor ($f_{\rm ap}$) for CO-VGS and xCOLD GASS galaxies.
            }
            \label{fig:int_rel}
        \end{figure*}

        The left panel of  Fig. \ref{fig:int_rel} shows the relation between $\rm CO(2-1)$ and $\rm CO(1-0)$ for the CO-VGS together with the xCOLD GASS comparison sample. For 15 CO-VGS galaxies, we obtained a detection in at least one line, so that we can calculate the mean value of the line ratio $R_{\rm 21}=I_{\rm CO(2-1)}/I_{\rm CO(1-0)}$ (listed in Table \ref{tab:int_rat}, together with the corresponding value for the xCOLD GASS sample). The mean values for the CO-VGS and xCOLD GASS samples are the same (within the errors). The mean values are not aperture-corrected, and therefore we have to take into account the different beam sizes of $\rm CO(1-0)$ and $\rm CO(2-1)$ when the ratios are interpreted.

        To interpret $R_{\rm 21}$ (Fig. \ref{fig:int_rel} right), we have to consider two parameters in addition to the excitation temperature of the gas: the source size relative to the beam, and the opacity of the molecular gas. For optically thick thermalised emission with a point-like distribution, we expect a ratio $R_{21}  = (\Theta_{\rm 1-0}/\Theta_{\rm 2-1})^2 = 4 $, with $\Theta_{\rm 1-0}$ and $\Theta_{\rm 2-1}$ being the FWHM of the $\rm CO(1-0)$ and $\rm CO(2-1)$ beam, respectively. On the other hand, for a source that is more extended than the beams, we expect $R_{21} \sim 0.6-1$ for optically thick gas in thermal equilibrium, where $R_{21}$ depends on the temperature of the gas, and $R_{21} > 1$ for optically thin gas.
 
        In order to better quantify the combined effect of sources size and intrinsic brightness temperature, we calculated the theoretical line ratio, $R_{\rm 21theo}$ (see Appendix \ref{sec:theo-line-ratio}) by modelling the CO emission with  the same 2D exponential distributions as  used for the aperture correction and adopting the IRAM 30 m telescope beam as a Gaussian function, with values for the full width at half maximum (FWHM) of  \makebox{$\Theta_{\rm 1-0} = 22 \,{\rm arcsec}$} and \makebox{$\Theta_{\rm 2-1} = 11\, {\rm arcsec}$}. We compare  the observed $ R_{\rm 21}$ empirical values with the theoretical $R_{\rm 21theo}$ values for different intrinsic brightness temperature ratios of the source \makebox{($\bar{T}_{\rm Bc2-1}/\bar{T}_{\rm Bc1-0}=1,\; 0.7$ and $0.5$)} in the right panel of Fig. \ref{fig:int_rel} .  For  optically thick gas in thermal equilibrium, an intrinsic brightness ratio of 0.6 corresponds to an excitation temperature of $\rm \sim5\,K$, 0.8 to $\rm \sim10\,K$, and 0.9 to $\sim21$ K; higher excitation temperatures yield a brightness temperature ratio $\sim1$ \citep{2009AJ....137.4670L}. The observed $ R_{\rm 21}$ in general follows the predicted trend of a decreasing value with $f_{\rm ap}$ well (which is an increasing function with galactic size). This indicates that the aperture correction we used is correct. For the void and comparison sample, the main part of the  values of $R_{\rm 21}$ lies below the line of $\bar{T}_{\rm Bc2-1}/\bar{T}_{\rm Bc1-0} = 0.7$, suggesting that the molecular gas is cold ($\rm < 10 \,K$). Interestingly, many galaxies have $\bar{T}_{\rm Bc2-1}/\bar{T}_{\rm Bc1-0} < 0.5,$ which might indicate the presence of low-density sub-thermally excited gas that is not in thermal equilibrium \citep{2009AJ....137.4670L, 2021MNRAS.504.3221D}.

        \begin{table*}
            \caption{$\rm CO(2-1)-to-CO(1-0)$ line ratio\label{tab:int_rat}}
            \centering
            \begin{tabular}{|c|ccccc|ccccc|ccccc|}     % 7 columns 
                \multicolumn{16}{c}{$R_{21}=I_{\rm CO(2-1)}/I_{\rm CO(1-0)}$}\\
                \hline
                Galaxies &  \multicolumn{5}{c|}{CO-VGS} &  \multicolumn{5}{c|}{CO-CS} &\multicolumn{5}{c|}{CO-VGS - CO-CS}\\
                considered  &  nº &  \multicolumn{3}{c}{mean} &  median &  nº &  \multicolumn{3}{c}{mean} &  median &\multicolumn{3}{c}{$\rm{ \Delta mean}$}&$\rm{ \sigma}$&KS\\
                (1)  &  (2) &  \multicolumn{3}{c}{(3)} &  (4) &  (5) &  \multicolumn{3}{c}{(6)} &  (7) &\multicolumn{3}{c}{(8)}&(9)&(10)\\
                \hline\hline 
                Upper lim. &      13/1 &     1.37 &         $\pm$ &         0.19 &       1.19 &     50/0 &    1.37 &        $\pm$ &        0.05 &      1.34 &       -0.00 &           $\pm$ &           0.19 &            -0.01 &        $1\times 10^{-3}$ \\
                \hline         
                Lower lim.   &      14/1 &     1.74 &         $\pm$ &         0.28 &       1.49 &     54/7 &    1.41 &        $\pm$ &        0.05 &      1.34 &       0.33 &           $\pm$ &           0.28 &             1.18 &       0.02 \\
                \hline
            \end{tabular}
            \tablefoot{ In the mean, only galaxies with an aperture correction of $f_{\rm ap} = 1.1-1.6$ were considered in order to make a consistent comparison. (1)  Upper row: Only detections and upper limits were taken into account. Lower row: Only detections and lower limits were taken into account. (2) n: Number of CO-VGS galaxies in the bin. $\rm n_{up}$: Number of upper limits of CO-VGS galaxies in the bin. (3) Mean emission line ratio of the CO-VGS galaxies in the bin taking lower limits into account. (4) Median emission line ratio of the CO-VGS galaxies in the bin. (5) - (7) The same for the CO-CS sample. (8) Difference of the mean emission line ratio between CO-VGS and CO-CS ($\rm{ \Delta mean}$) and its error ($\rm err( \Delta mean)$). (9) $\rm \sigma = { \Delta mean}/err({\Delta mean})$. (10) p-value of the Kolmogorov-Smirnov test.
            }
        \end{table*}
   
\section{Discussion}
    \label{sec:discussion}
 
    The molecular gas masses for 20 objects presented in this paper are the largest sample of void galaxies with molecular gas data so far. This enables us to statistically compare the properties of void galaxies to those in filaments and walls. 
 
    Our results show no significant difference in the mean $M_{\rm H_2}$ and $M_{\rm H_2}/M_\star$ for different mass bins compared to the comparison sample. The exception is the intermediate stellar mass bin, especially in the SF sub-sample,  where $M_{\rm H_2}$ in void galaxies might be lower than for galaxies in filaments and walls. Our mean value for $M_{\rm H_2}/M_\star$ of  SF sample ($\log_{10}M_{\rm HI}/M_\star=-1.2$) agrees with the values found by \citet{2021arXiv210104389C} for field and filament galaxies ($\log_{10}M_{\rm HI}/M_\star=-1.3$).
 
    For the SFE, the CO-VGS and control sample also agree well. The SFE is lower for the CO-VGS in the lowest-mass bin, but the number of objects is small in this bin. The SFE of four VGS galaxies with stellar masses just below  $10^{9.0} {\rm M_\odot}$ agrees better with the mean SFE of the comparison sample for the lowest stellar mass bin ($M_\star$ between $10^{9.0}$ and $\rm 10^{9.5} M_\odot$). This indicates that the low mean SFE that we find for the CO-VGS in the low stellar mass bin needs to be confirmed for a larger sample of void galaxies before any firm conclusions can be drawn.
 
    The atomic gas mass fraction in the void galaxies follows the trend of the control sample for  $M_\star<10^{9.0}{\rm M_\odot}$ quite well and has lower values for higher $M_\star$, drawing a steeper trend for void galaxies with $M_\star>10^{9.0}\, {\rm M_\odot}$. This agrees with \cite{2012AJ....144...16K}, who found evidence for a slight lack of $M_{\rm H{\scriptscriptstyle  I}}$ for $M_\star \gtrsim 10^{9.0}\, {\rm M_\odot}$  in void galaxies (for the same void galaxies as in our study, but for a different control sample). In contrast, \cite{2021ApJ...906...97F} found  a small enhancement of $M_{\rm H{\scriptscriptstyle  I}}$ in void galaxies (up to $\sim$ 0.2 dex), especially for galaxies with $M_\star < 10^{9.5}\, {\rm M_\odot}$, for a sample of $\sim 900$ void galaxies and a control sample of $\sim 8500$ galaxies. 
 
    \cite{2021arXiv210104389C} found for late-type galaxies that the atomic gas mass fraction decreases with the local density on average from field ($\log_{10}M_{\rm HI}/M_\star=-0.47$) and filaments ($\log_{10}M_{\rm HI}/M_\star=-0.52$) to clusters ($\log_{10}M_{\rm HI}/M_\star=-1.10$), which means that galaxies might be stripped of their gas while falling from field and filaments into clusters, or they might be affected by tidal interactions \citep{2021arXiv211007836C}. We find similar average values in filaments ($\log_{10}M_{\rm HI}/M_\star=-0.50$) and slightly higher values in voids ($\log_{10}M_{\rm HI}/M_\star=-0.30$ for SF galaxies), where the density is lower \citep{2012AJ....144...16K,2012MNRAS.421..926P}, following the same trend. However, we find the opposite result for massive galaxies ($M_\star > 10^{9.5}\, {\rm M_\odot}$), where the atomic gas mass ratio is lower in voids ($\sim3\sigma$). If this discrepancy is confirmed for a larger number of void galaxies, it suggests that (because gas-stripping processes, such as ram pressure or frequent interactions, are unusual in voids) the lower atomic gas mass in massive void galaxies might be due to a gas deficiency in the inter-galactic medium of voids, or that slower gas accretion processes take place in void galaxies.
 
    The mean molecular-to-atomic gas mass ratio is consistent with  that of the control sample, except for the  highest stellar mass bin ($M_\star >{\rm 10^{10.0}\,M_\odot}$).  As we do not find differences with the mass of the molecular gas between samples, the  result for the highest-mass bin  seems to be driven by the lower atomic gas mass of high-mass void galaxies with respect to the control sample. It may also be driven by the low number of VGS galaxies (three) in this stellar mass bin, however.
  
    The mean sSFR values of the VGS are very close to the mean values of the CCS for the entire sample and for the SF sample. The mean value of the VGS is up to $|\sigma| \sim 3$ below that of the CCS for one individual mass bin, but no trends with stellar mass are visible. We thus do not find evidence for a general significant difference of the sSFR between the void and the control sample, and in particular, we do not find any evidence at all for an enhancement of the sSFR.

    When we compare our results to those from the literature, we find that a number of other studies found no differences in the sSFR of void galaxies either \citep{2006MNRAS.372.1710P,2012AJ....144...16K, 2014MNRAS.445.4045R}. Others found that voids are populated by galaxies with higher sSFR \citep{2005ApJ...624..571R, 2016MNRAS.458..394B, 2021ApJ...906...97F}, however. The direct comparison is not straightforward, however, because the sample environment might playd a role; for instance, \cite{2016MNRAS.458..394B} only use field and isolated galaxies for their comparison sample, but in the present work, we used the xCOLD GASS sample, which is a representative sample of SDSS galaxies in filaments and walls, after removing galaxies inhabiting voids or clusters. Furthermore, there seems to be a clear dependence on the SFR tracer that is used. In Appendix \ref{sec:SFR} we show a comparison between the different SFR tracers we used  for our control samples: $\rm H\alpha$ maps were used for the VGS galaxies \citep{2016MNRAS.458..394B} and the HI-CALIFA sample \citep{2015A&A...584A..87C}, whereas the SFR of the xCOLD GASS sample was derived from near-ultraviolet (NUV) and mid-infrared (MIR) emission \citep{2017ApJS..233...22S}. In addition, the SFR from the MPA-JHU is frequently used in the literature  \citep{2006MNRAS.372.1710P,2012AJ....144...16K, 2014MNRAS.445.4045R, 2005ApJ...624..571R} and is available for the VGS and the control samples. Our comparison shows that the MPA-JHU SFR  systematically gives higher SFRs for the void galaxies compared to the other methods, and that the effect increases for lower  SFRs (Fig. \ref{fig:sfrmpacomp}). The comparison between the MPA-JHU and other SFR tracers for the xCOLD GASS and EDGE-CALIFA galaxies shows that this trend continues to higher SFRs and suggests that the MPA-JHU  progressively underestimates the SFRs with increasing SFR. These results might reflect a problem in the aperture correction as MPA-JHU seems to overestimate the SFR for compact galaxies such as the VGS (values of $R_{\rm 90} \sim 4-15 \,{\rm arcsec}$) and underestimates the SFR for larger galaxies such the EDGE-CALIFA galaxies ($R_{\rm 90} \sim 20-45\,{\rm arcsec}$;  see also Fig. \ref{fig:sfrr90comp}, where we compare the different SFR tracers as a function of  the apparent size of the galaxies). The SFR tracers used in our comparison are more robust (see Fig. \ref{fig:comp-SFR_Janow}). Thus, the use of different SFR tracers might explain the disagreement of our result compared to \citet{2006MNRAS.372.1710P,2012AJ....144...16K, 2014MNRAS.445.4045R} and \citet{2005ApJ...624..571R}, who  used the MPA-JHU SFR.
 
    All this makes it difficult to draw any strong conclusion about the apparent disagreement with previous works, but it indicates that a revision of the subject is required that takes a careful look at the comparison sample and the SFR tracer used for the comparison. This is beyond the scope of the present paper, especially because we still lack enough number statistics to carry out a more detailed study. This might be one of the scopes of CO-CAVITY, which will enhance the statistics.
 
    There is no numerical prediction about the molecular gas content of void galaxies. Our finding of similar  molecular gas masses or molecular gas mass fractions between void galaxies and the comparison sample is a clear constraint for future simulations of galaxy evolution in voids. Some numerical simulations \citep{2011ApJ...741...99C} predict that the cold-gas inflow rate at redshift $\rm z=0$ will be higher for void than for cluster galaxies, even more so in the low-mass range, but there are no predictions about the colder star-forming phase. These simulations predict a clearly higher sSFR for void galaxies with masses  $10^9.0\,{\rm M_\odot}<M_\star<10^{10.0}\,{\rm M_\odot}$ and only marginally higher for a higher mass range. Again, it is not straightforward to compare this prediction with our results, not only due to the low number statistics, but also because the simulations compare void galaxies with cluster galaxies and our comparison sample includes  non-void environments and no cluster galaxies.
 
\section{Conclusions}
    \label{sec:conclusions}

    We have observed the $\rm CO(1-0)$ and $\rm CO(2-1)$ emission lines of 20 void galaxies from the VGS with the IRAM 30 m telescope. The $\rm CO(1-0)$ line was detected for 13 galaxies and the $\rm CO(2-1)$ for 14 galaxies, allowing us to derive the molecular gas mass for 15 detected galaxies and calculate upper limits for 5 non-detected galaxies. This represents the largest CO sample of void galaxies up to date.  

    We selected a  comparison sample from the xCOLD GASS and EDGE-CALIFA samples, which have available data for stellar mass, star formation rate, atomic gas mass, and molecular gas mass. Most of the VGS galaxies are star-forming main-sequence galaxies, but the control sample has many quiesicent galaxies with a low sSFR. To take this into account, we defined  star-forming sub-samples for the VGS and the control sample by selection galaxies close to the star-forming main sequence and carried out the entire analysis for these sub-samples  as well. Based on these data and samples, we studied the specific star formation rate, the molecular gas mass, the molecular gas mass fraction, the star formation efficiency, the atomic gas mass fraction, and the molecular-to-atomic gas mass ratio  by comparing the  mean values of the void galaxies in different stellar mass bins to those of the control  samples. The main conclusions are listed below.

    \begin{enumerate}
        \item We do not find any clear difference for the molecular gas mass or molecular gas mass fraction between void galaxies and the comparison sample. Void galaxies seem to have the same molecular gas fraction as galaxies in filaments and walls.
 
        \item We did not find any  evidence for differences in the SFE, except for the lowest-mass bin ($10^{9.0}{\rm M_{\odot}}\leq M_\star<10^{9.5}{\rm M_{\odot}}$), in which the SFE of void galaxies is significantly ($|\sigma| > 3$) below that of the control sample. However, due to the low number of galaxies in this sub-sample (four to five objects), the results need to be confirmed for a larger sample.
 
        \item There is some evidence for a lower atomic gas mass fraction and a higher molecular-to-atomic gas mass ratio in void galaxies for  $M_\star> 10^{9.5}$ and $M_\star> 10^{10.0}{\rm M_{\odot}}$, respectively. The mean values for lower stellar masses are the same as for the control sample within the errors. Again, the results for the higher stellar masses need to be confirmed for larger sample because they are derived from a low number of galaxies (three to five objects).
 
        \item We do not find any clear difference in the sSFR  between void galaxies and the control sample, and in particular, we do not find an enhancement for void galaxies. 
 
        \item The $\rm CO(2-1)$-to-$\rm CO(1-0)$ line ratio does not show any clear difference between void galaxies and the control sample. 
    \end{enumerate}

    Our study was based on a small number of galaxies, and some of our conclusions are based on low number statistics. CO-CAVITY, together with CAVITY, plans to overcome this limitation by providing observational data of the star formation and ionized and neutral gas for a sample of several hundred  void galaxies.

\begin{acknowledgements}

    We appreciate the detailed and useful comments of the referee which helped to improve the clarity of the paper. We acknowledge support (proyect AYA2017-84897-P) from the Spanish Ministerio de Econom\'\i a y Competitividad, from the European Regional Development Funds (FEDER) and the Junta de Andaluc\'ia (Spain) grants FQM108. This work is based on observations carried out under project numbers  075-19 with the IRAM 30-m telescope. IRAM is supported by INSU/CNRS (France), MPG (Germany) and IGN (Spain).
    
    J.~F-B  acknowledges support through the RAVET project by the grant PID2019-107427GB-C32 from the Spanish Ministry of Science, Innovation and Universities (MCIU), and through the IAC project TRACES which is partially supported through the state budget and the regional budget of the Consejer\'ia de Econom\'ia, Industria, Comercio y Conocimiento of the Canary Islands Autonomous Community.

    SDP is grateful to the Fonds de Recherche du Qu\'ebec - Nature et Technologies and acknowledge financial support from the Spanish Ministerio de Econom\'ia y Competitividad under grants AYA2016-79724-C4-4-P and PID2019-107408GB-C44, from Junta de Andaluc\'ia Excellence Project P18-FR-2664, and also acknowledge support from the State Agency for Research of the Spanish MCIU through the `Center of Excellence Severo Ochoa' award for the Instituto de Astrof\'isica de Andaluc\'ia (SEV-2017-0709).
    
    This research made use of Astropy, a community-developed core Python (http://www.python.org) package for Astronomy \citep{2013A&A...558A..33A, 2018AJ....156..123A}; ipython \citep{PER-GRA:2007}; matplotlib \citep{Hunter:2007}; SciPy, a collection of open source software for scientific computing in Python \citep{2020SciPy-NMeth}; APLpy, an open-source plotting package for Python \citep{2012ascl.soft08017R,2019zndo...2567476R}; and NumPy, a structure for efficient numerical computation \citep{2011CSE....13b..22V}.
\end{acknowledgements}

\bibliographystyle{bibtex/aa}
\bibliography{biblio}

\clearpage

\begin{appendices}
\counterwithin{figure}{section}

\section{Theoretical $\rm CO(2-1)$-to-$\rm CO(1-0)$ line ratio }
    \label{sec:theo-line-ratio}

    In this section we estimate the theoretical value of the $\rm CO(2-1)$-to-$\rm CO(1-0)$ line ratio, $R_{\rm 21theo}$, assuming a Gaussian power pattern of the antenna and an exponential distribution of the CO emission in the galaxy.
 
    The measured main beam temperature can be expressed as
    \begin{equation} \label{eq:tmbdef}
        T_{\rm  mb}(\nu)=\frac{\int_{-\infty}^{\infty}\int_{-\infty}^{\infty}T_{\rm  B}(x,y,\nu)P_{\rm  n}(x,y){\rm d}x{\rm  d}y}{\int_{-\infty}^{\infty}\int_{-\infty}^{\infty}P_{\rm n}(x,y){\rm d}x{\rm  d}y},
    \end{equation}
    where $P_{\rm n}$ is the normalised power pattern of the antenna beam, and $T_{\rm  B}$ is the brightness temperature of the source.

    $P_{\rm n}$ is assumed to be a Gaussian distribution, $G[\Theta]$,
  
    \begin{equation} 
        \label{eq:pn}
        P_{\rm n}(x,y)= \exp{\left(-\ln2 \left[\left(\frac{2x}{\Theta}\right)^2+\left(\frac{2y}{\Theta}\right)^2\right]\right)}= G[\Theta]
    ,\end{equation}
    where $\rm \Theta$ is the FWHM of the antenna beam ($\rm \Theta_{10}=22''$ and $\rm \Theta_{21}=11''$).

    The velocity-integrated emission line intensity is

    \begin{equation} \label{eq:icodef}
        I_{\rm CO} =\int_{line}^{}T_{\rm mb}(\nu){\rm d}\nu.
    \end{equation}
   
    In the same way as for the aperture correction, we assumed that the spatial distribution of the intrinsic brightness temperature is an exponential disc with an inclination $i$ with respect to the line of sight. This intrinsic brightness temperature distribution, as observed in the coordinate system ($x,y$) in the plane of the sky, for this case is

    \begin{equation} \label{eq:tb}
        { T_{\rm  B}(x,y,\nu)= \frac{T_{\rm  Bc}(\nu)}{\cos(i)} \exp \left( -\frac{\sqrt{x^2+(y/\cos(i))^2}}{r_{\rm e}} \right)=T_{\rm  Bc}(\nu) E[r_{\rm e}, i]}
    ,\end{equation}
    where $T_{\rm Bc}(\nu)$ is the intrinsic brightness temperature at the centre of the source, $r_{\rm e}$ is the exponential scale factor, which is derived from $R_{\rm 90}$ as described in Sect.~\ref{sec:fap}, and $i$ is the inclination of the galaxy. The factor involving the inclination appears twice in the function: Firstly, in the denominator, because the intrinsic brightness temperature is higher in an inclined galaxy by $1/\cos(i)$ for a given distance on the sky, ($x,y$), because of the higher apparent disc-thickness, and secondly, because the physical position along the radius of the galactic disc, $y^\prime$, is higher than the projected position on the sky, $y$.
   
    Inserting this distribution (eq.~\ref{eq:tb}) and the Gaussian beam distribution (eq.~\ref{eq:pn}) into eq.~\ref{eq:tmbdef}, we obtain for the measured main-beam temperature at the central position of the galaxy
   
    \begin{equation}
        \small
        \begin{split} 
            & T_{\rm  mb}(\nu)  =  \frac{\int_{-\infty}^{\infty}\int_{-\infty}^{\infty} T_{\rm  Bc}(\nu) E[r_{\rm e}, i] G[\Theta]  {\rm d}x{\rm d}y }{\int_{-\infty}^{\infty}\int_{-\infty}^{\infty}G[\Theta]{\rm d}x{\rm  d}y} =\\
            & \frac{\int_{-\infty}^{\infty}\int_{-\infty}^{\infty} \frac{T_{\rm  Bc}(\nu)}{\cos(i)}  \exp \left( -\frac{\sqrt{x^2+\left(\frac{y}{\cos(i)}\right)^2}}{r_{\rm e}} \right) \exp{\left(-\ln2 \left[\left(\frac{2x}{\Theta}\right)^2+\left(\frac{2y}{\Theta}\right)^2\right]\right)} {\rm d}x{\rm d}y}{\int_{-\infty}^{\infty}\int_{-\infty}^{\infty}G[\Theta]{\rm d}x{\rm d}y} =\\ & \frac{\int_{-\infty}^{\infty}\int_{-\infty}^{\infty} T_{\rm  Bc}(\nu) \exp \left( -\frac{\sqrt{x^2+\left(y^\prime\right)^2}}{r_{\rm e}} \right) \exp{\left(-\ln2 \left[\left(\frac{2x}{\Theta}\right)^2+\left(\frac{2y^\prime \cos(i)}{\Theta}\right)^2\right]\right)} {\rm d}x{\rm  d}y^\prime}{\int_{-\infty}^{\infty}\int_{-\infty}^{\infty}G[\Theta]{\rm d}x{\rm  d}y}=\\
            & \frac{\int_{-\infty}^{\infty}\int_{-\infty}^{\infty} T_{\rm  Bc}(\nu)E^\prime[r_{\rm e}] G^\prime[\Theta, i]  {\rm d}x{\rm  d}y^\prime}{\int_{-\infty}^{\infty}\int_{-\infty}^{\infty}G[\Theta]{\rm d}x{\rm  d}y}
        \end{split}
    ,\end{equation}

    where  we made the substitution $y^\prime = y/\cos(i)$.

    We suppose that the line shape, $\psi(\nu)$, is the same for both $\rm CO(1-0)$ and $\rm CO(2-1)$ emission lines,
   
    \begin{equation} \label{eq:colinerat1}
        \begin{split}
            \frac{T_{\rm Bc2-1}(\nu)}{T_{\rm Bc1-0}(\nu)}=\frac{\bar{T}_{\rm Bc2-1}\psi(\nu)}{\bar{T}_{\rm Bc1-0}\psi(\nu)}=\frac{\bar{T}_{\rm Bc2-1}}{\bar{T}_{\rm Bc1-0}}
        \end{split}.
    \end{equation}
 
    Then,  $R_{rm 21theo}$ is calculated as  
    \begin{equation} \label{eq:colinerat2}
        \begin{split}
            R_{\rm 21theo}=\frac{I_{\rm CO(2-1)}}{I_{\rm CO(1-0)}} = \frac{\bar{T}_{\rm Bc2-1}}{\bar{T}_{\rm Bc1-0}}\left(\frac{\Theta_{1-0}}{\Theta_{2-1}}\right)^2 \frac{\int_{-\infty}^{\infty} \int_{-\infty}^{\infty}E^\prime[r_{\rm e}]G^\prime[\Theta_{2-1},i]{\rm d}x{\rm  d}y^\prime}{\int_{-\infty}^{\infty} \int_{-\infty}^{\infty}E^\prime[r_{\rm e}]G^\prime[\Theta_{1-0},i]{\rm d}x{\rm  d}y^\prime}
        \end{split}
    ,\end{equation}   
    where we used the identity
    $\int_{-\infty}^{\infty}\int_{-\infty}^{\infty}G[\Theta]{\rm d}x{\rm  d}y =\frac{\pi}{4\ln2} \Theta^2$.
    The integrals were calculated numerically for each galaxy in the present paper.

\section{Selection of SFR tracer}
    \label{sec:SFR}

    For a correct comparison of the different samples, we need to employ the same tracer for the SFR, or, if this is not possible, test whether different tracers give consistent results. In addition, the calibrations need to be based on the same initial mass function (IMF). Here, we used the Kroupa IMF \citep{2001MNRAS.322..231K}, which is very similar to the IMF by Chabrier \citep{2003PASP..115..763C}.

    The SFR of the xCOLD GASS sample was derived from the NUV and MIR emission and  followed the prescription from \cite{2017MNRAS.466.4795J}. The SFR of the HI-CALIFA sample was derived and calibrated from extinction-corrected $\rm H\alpha$ fluxes in \cite{2015A&A...584A..87C}, resulting in practically the same prescription as we used (eq.~\ref{eq:sfr}) (difference $<$ 2\%, which is negligible). 

    Finally, the SFRs from the MPA-JHU were available for all samples.

    \begin{figure*}[!h]
        \centering
        \includegraphics[width=0.8\hsize]{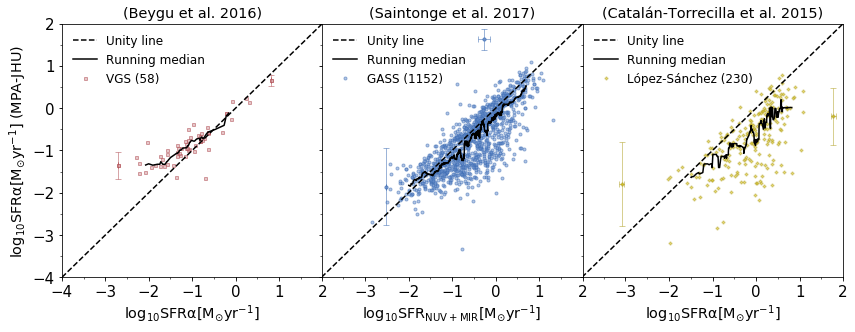}
        \caption{Star formation rate
from the MPA-JHU compared to the SFR derived from different tracers. Left: SFR of the VGS derived from $\rm H\alpha$ maps \citep{2016MNRAS.458..394B}. Centre: SFR of the xCOLD GASS sample derived from NUV and MIR emission \citep{2017ApJS..233...22S}. Right: SFR of the HI-CALIFA sample derived from $\rm H\alpha$ maps \citep{2015A&A...584A..87C}.\label{fig:sfrmpacomp}
        }
    \end{figure*}
    
    \begin{figure*}[!h]
        \centering
        \includegraphics[width=0.8\hsize]{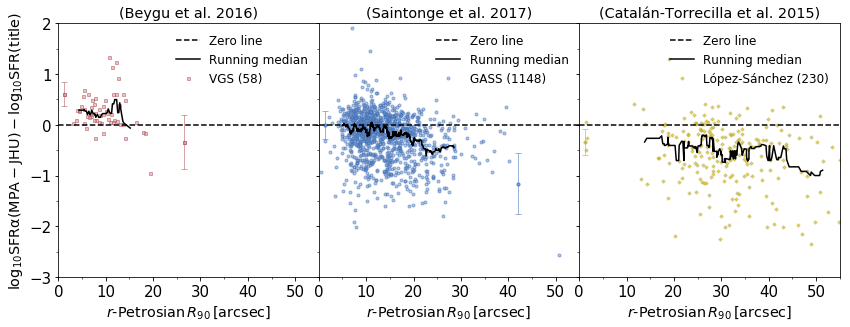}
        \caption{Difference between the SFR from the MPA-JHU and the SFR derived from different tracers represented as a function of the apparent size of the galaxy ($r$-Petrosian $R_{\rm 90}$). Left: Difference between the SFR of the VGS derived from the MPA-JHU and the SFR derived from $\rm H\alpha$ maps \citep{2016MNRAS.458..394B}. Centre: Difference between the SFR of the xCOLD GASS sample derived from the MPA-JHU and the SFR derived from NUV and MIR emission \citep{2017ApJS..233...22S}. Right: Difference between the SFR of the HI-CALIFA sample derived from the MPA-JHU and the SFR derived from $\rm H\alpha$ maps \citep{2015A&A...584A..87C}.\label{fig:sfrr90comp}
        }
    \end{figure*}
    
    \begin{figure*}[!h]
        \centering
        \includegraphics[width=0.8\hsize]{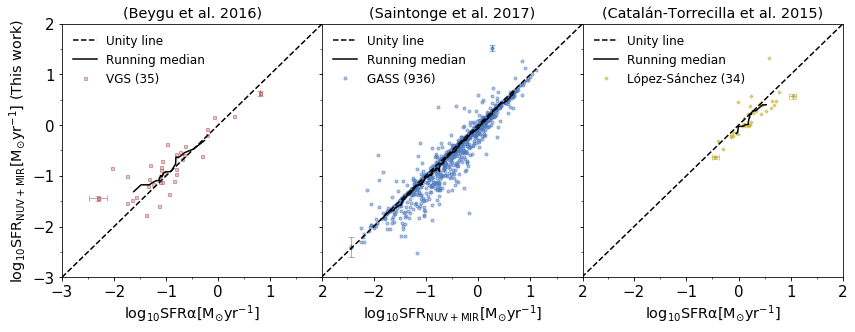}
        \caption{Star formation rate tracer derived from NUV and MIR emission, following the prescription in  \cite{2017MNRAS.466.4795J}, derived in the present paper and compared with different SFR tracer. Left: SFR of the VGS derived from $\rm H\alpha$ maps \citep{2016MNRAS.458..394B}. Centre: SFR of the xCOLD GASS sample derived from NUV and MIR emission \citep{2017ApJS..233...22S}. Right: SFR of the HI-CALIFA sample derived from $\rm H\alpha$ maps \citep{2015A&A...584A..87C}.\label{fig:comp-SFR_Janow}
        }
    \end{figure*}
    
    In Fig. \ref{fig:sfrmpacomp} we compare the SFR from the MPA-JHU \citep{2004MNRAS.351.1151B} to the SFR derived from different tracers in the different studies. We find that for galaxies with a low SFR ($\rm <10^{-1}\,M_\odot\,yr^{-1}$), the SFR from the MPA-JHU is overestimated for the VGS on average, but it is underestimated for the comparison sample. A similar result was obtained by \cite{2017A&A...599A..71D}. This suggests that the $\rm H\alpha$ aperture correction of the MPA-JHU might overestimate the SFR for small galaxies such as the VGS (values of $R_{\rm 90} \sim 4-15 \,{\rm arcsec}$) and might underestimate it for larger galaxies such as the HI-CALIFA sample or xCOLD GASS ($R_{\rm 90} \sim 20-45\, {\rm arcsec}$). Fig. \ref{fig:sfrr90comp} shows the SFR differences between the MPA-JHU and the tracer used in this work (for the different samples) as a function of the apparent size of the galaxy.   
   
    In Fig.~\ref{fig:comp-SFR_Janow} we compare the SFR derived from NUV and MIR to the values derived from $\rm H\alpha$ in order to test whether we obtain consistent results with the different prescription.  First, we repeated the calculation of the SFR derived from NUV and MIR following \cite{2017MNRAS.466.4795J} for the xCOLD GASS and compare their results to ours in Fig. \ref{fig:comp-SFR_Janow} (centre). We conclude that we have repeated the method accurately. Then, we applied this method to the VGS and the HI-CALFIA samples and compare in Fig. \ref{fig:comp-SFR_Janow} (left and right) our results to the SFR derived from $\rm H\alpha$ maps by \cite{2016MNRAS.458..394B} and \cite{2015A&A...584A..87C}, respectively.
    
    The comparisons show some scatter, but the running median follows the unity line for the VGS, the xCOLD GASS, and HI-CALIFA samples very well. We conclude that the SFR tracer from \cite{2017MNRAS.466.4795J} is repeated accurately in the present paper, and it is equivalent to the SFR tracers derived from $\rm H\alpha$ maps in \cite{2016MNRAS.458..394B} and \cite{2015A&A...584A..87C}.
    
    Finally, we used the $\rm SFR$ derived from $\rm H_{\alpha}$ maps by \cite{2016MNRAS.458..394B} for the VGS, the $\rm SFR$ derived from $\rm H\alpha$ maps in \cite{2015A&A...584A..87C} for the HI-CALIFA sample, and the $\rm SFR$ derived from NUV and MIR emission by \cite{2017ApJS..233...22S} for the xCOLD GASS sample.
    
\FloatBarrier
   
\section{CO emission line spectra}
    \label{sec:cospectra}
 
\begin{figure*}
        \centering
        \begin{tabular}{ccc}
            \centering
            \includegraphics[width=5cm]{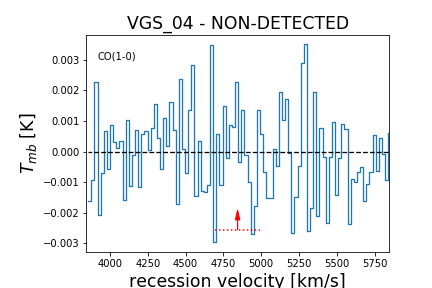} &
            \includegraphics[width=5cm]{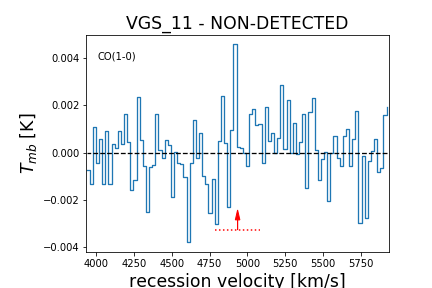} &
            \includegraphics[width=5cm]{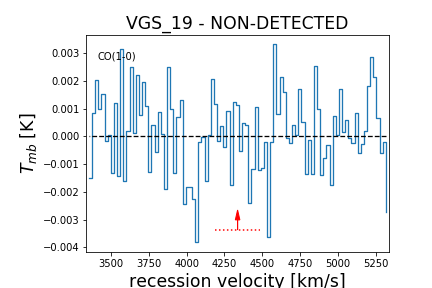} \\
            \includegraphics[width=5cm]{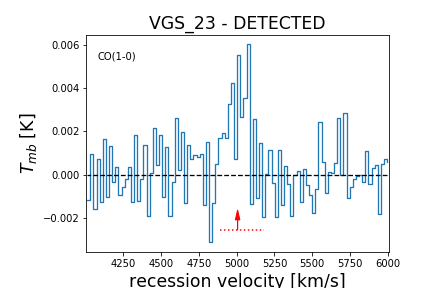} &
            \includegraphics[width=5cm]{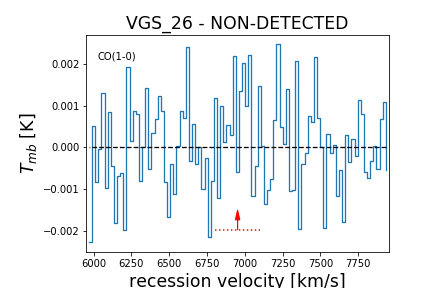} &
            \includegraphics[width=5cm]{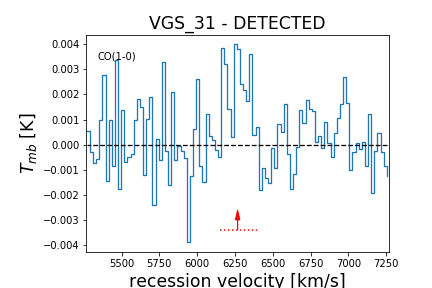} \\
            \includegraphics[width=5cm]{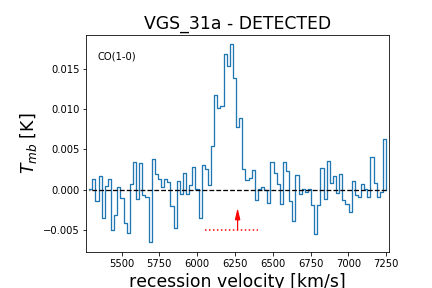} &
            \includegraphics[width=5cm]{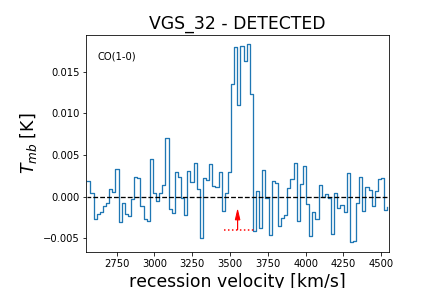} &
            \includegraphics[width=5cm]{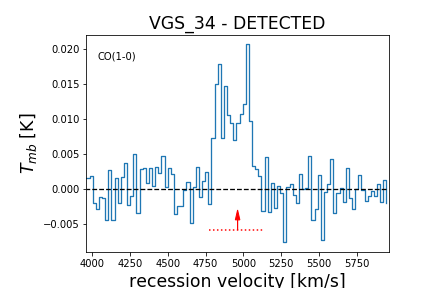} \\
            \includegraphics[width=5cm]{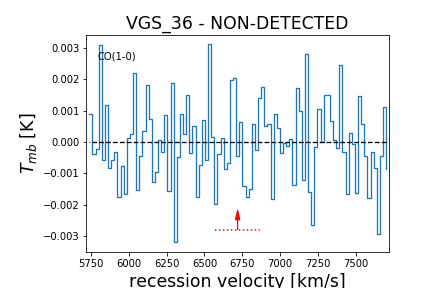} &
            \includegraphics[width=5cm]{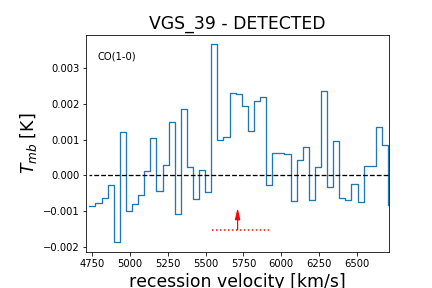} &
            \includegraphics[width=5cm]{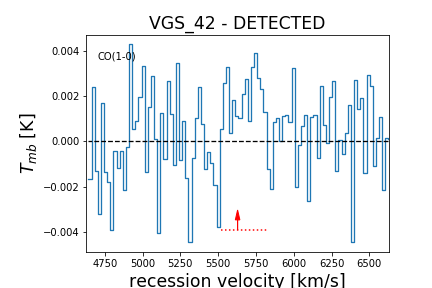} \\
            \includegraphics[width=5cm]{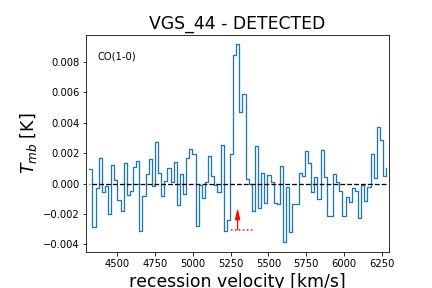} &
            \includegraphics[width=5cm]{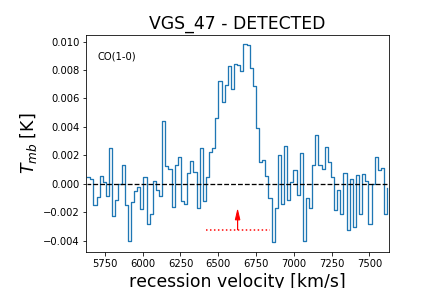} &
            \includegraphics[width=5cm]{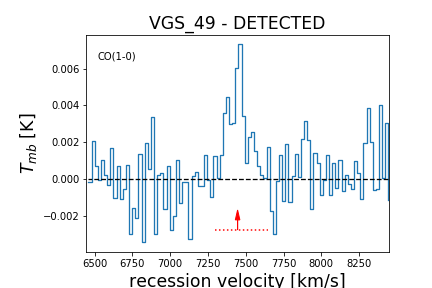} \\
            \includegraphics[width=5cm]{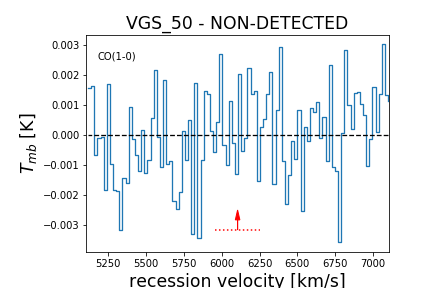} &
            \includegraphics[width=5cm]{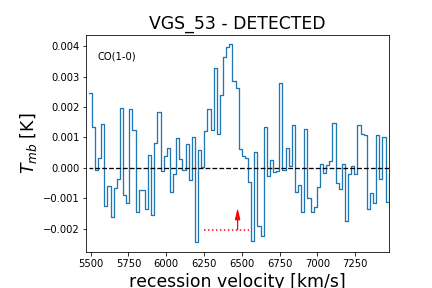} &
            \includegraphics[width=5cm]{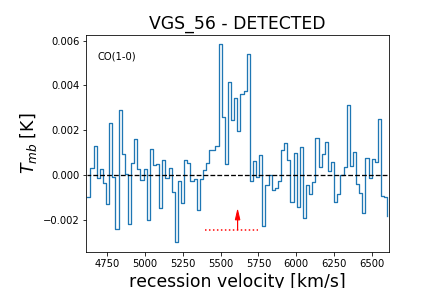} \\
            \includegraphics[width=5cm]{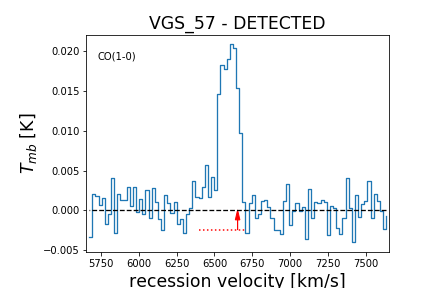} &
            \includegraphics[width=5cm]{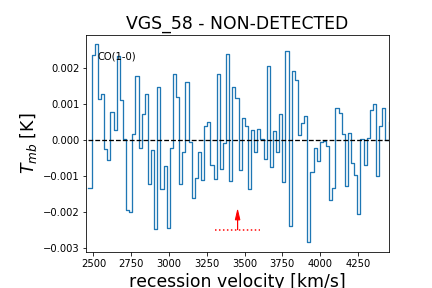} \\
        \end{tabular}
        \caption{Spectral representation of the $\rm CO(1-0)$ emission line $T_{\rm mb}$ in K at $\rm \sim20\,km\,s^{-1}$ of the velocity resolution. The red arrow indicates the optical heliocentric recession velocity. The dotted red line shows the zero-level line width at which the velocity-integrated intensity has been calculated. \label{fig:spec10}
        }
    \end{figure*}

\begin{figure*}
        \centering
        \begin{tabular}{ccc}
            \centering
            \includegraphics[width=5cm]{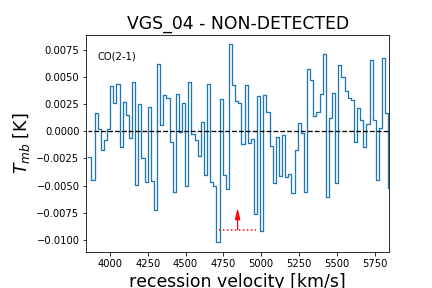} &
            \includegraphics[width=5cm]{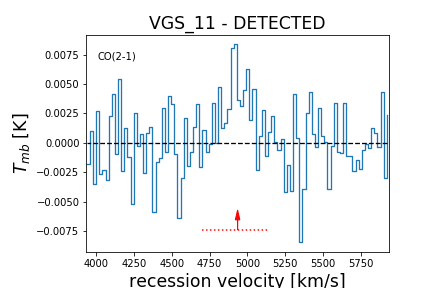} &
            \includegraphics[width=5cm]{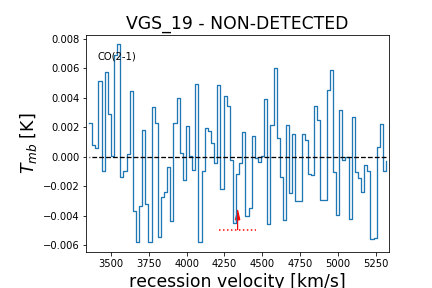} \\
            \includegraphics[width=5cm]{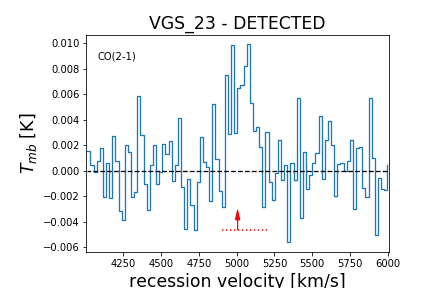} &
            \includegraphics[width=5cm]{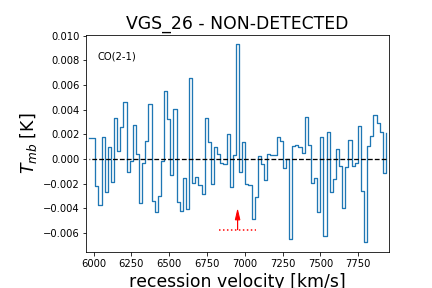} &
            \includegraphics[width=5cm]{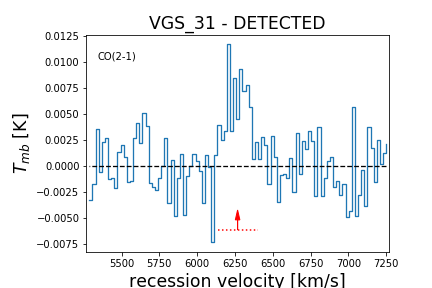} \\
            \includegraphics[width=5cm]{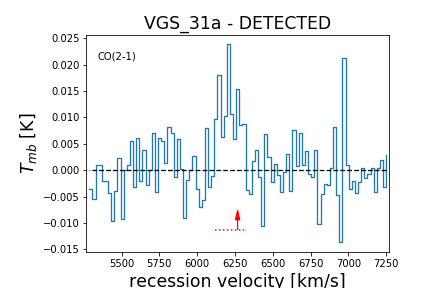} &
            \includegraphics[width=5cm]{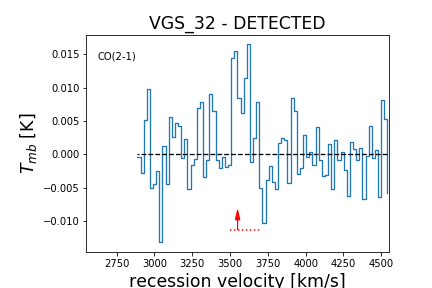} &
            \includegraphics[width=5cm]{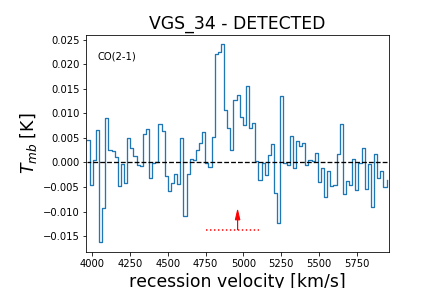} \\
            \includegraphics[width=5cm]{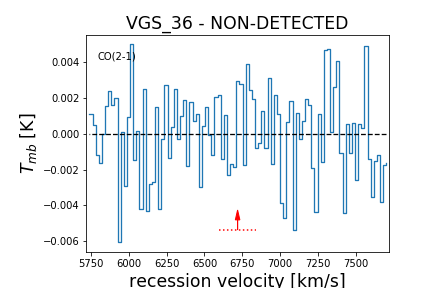} &
            \includegraphics[width=5cm]{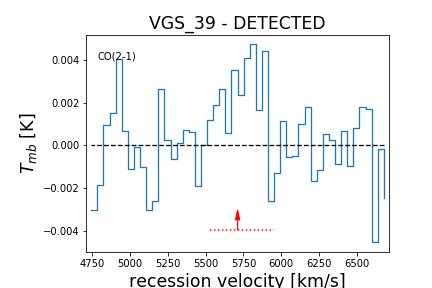} &
            \includegraphics[width=5cm]{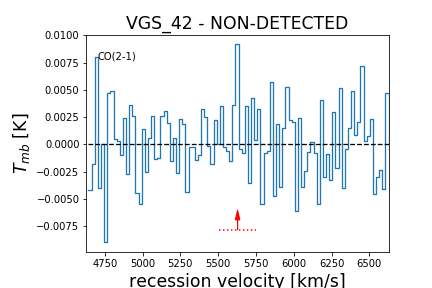} \\
            \includegraphics[width=5cm]{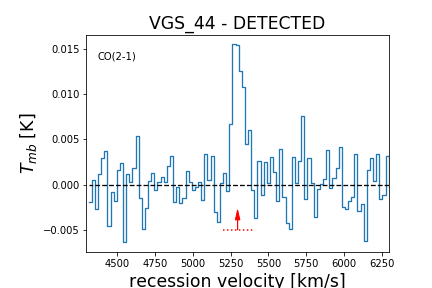} &
            \includegraphics[width=5cm]{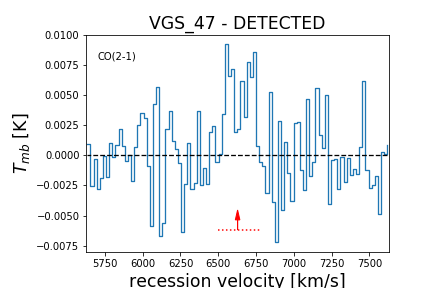} &
            \includegraphics[width=5cm]{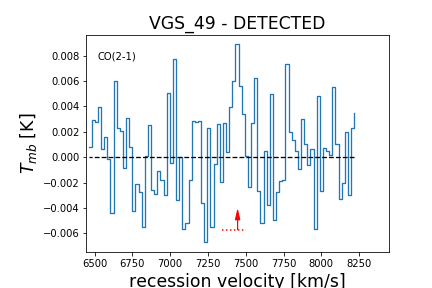} \\
            \includegraphics[width=5cm]{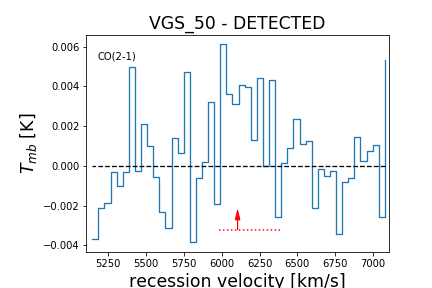} &
            \includegraphics[width=5cm]{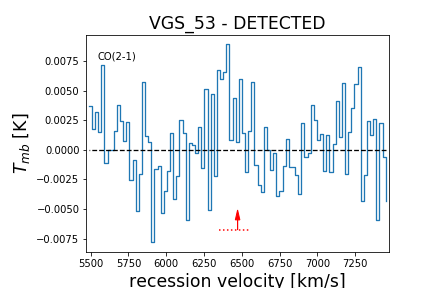} &
            \includegraphics[width=5cm]{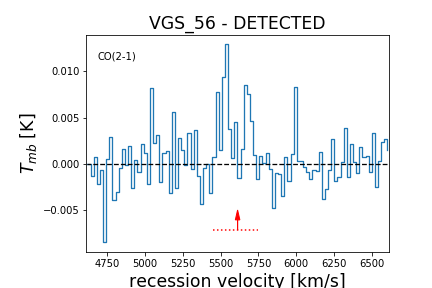} \\
            \includegraphics[width=5cm]{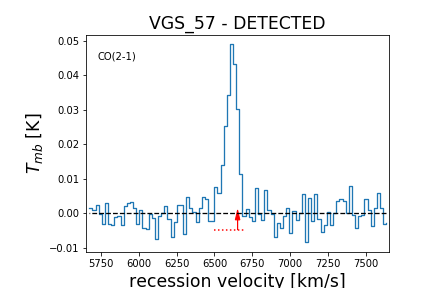} &
            \includegraphics[width=5cm]{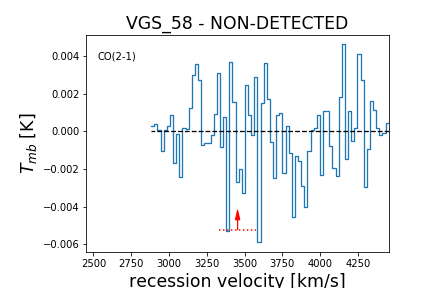} \\
        \end{tabular}
        \caption{Spectral representation of the $\rm CO(2-1)$ emission line $T_{\rm mb}$ in K at $\rm \sim20\,km\,s^{-1}$ of the velocity resolution. The red arrow indicates the optical heliocentric recession velocity. The dotted red line shows the zero-level line width at which the velocity-integrated intensity has been calculated.\label{fig:spec21}
        }
    \end{figure*}

    Figs. \ref{fig:spec10} and \ref{fig:spec21} show the observed spectra of the $\rm CO(1-0)$ and $\rm CO(2-1)$ emission lines, respectively.

\end{appendices}
\end{document}